\font \eightrm=cmr8

\documentclass[12pt]{amsart}
\usepackage{amssymb}
\usepackage{amsmath}
\usepackage{amsthm}
\usepackage{amscd}
\usepackage[all]{xy}               
\usepackage{epsfig,exscale}
\usepackage{color}
\usepackage{colordvi}

\newtheorem{thm}{Theorem}
\newtheorem{cor}[thm]{Corollary}

\newtheorem{lem}[thm]{Lemma}
\newtheorem{prop}[thm]{Proposition}

\newtheorem{problem}[thm]{Problem}

\def\BCH{{\rm{BCH}}}

\newcommand{\nc}{\newcommand}
\nc{\mrm}[1]{{\rm #1}}
\nc{\dirlim}{\displaystyle{\lim_{\longrightarrow}}\,}
\nc{\invlim}{\displaystyle{\lim_{\longleftarrow}}\,}
\nc{\vep}{\varepsilon}
\nc{\ep}{\epsilon}
\nc{\delete}[1]{{}}

\nc{\mchar}{\mrm{Char}}
\nc{\Hom}{\mrm{Hom}}
\nc{\id}{\mrm{id}}

\nc{\remark}{\noindent{\bf{Remark:}}}
\nc{\remarks}{\noindent{\bf{Remarks:}}}

\nc{\BA}{{\Bbb A}}
\nc{\CC}{{\Bbb C}}
\nc{\DD}{{\Bbb D}}
\nc{\EE}{{\Bbb E}}
\nc{\FF}{{\Bbb F}}
\nc{\GG}{{\Bbb G}}
\nc{\HH}{{\Bbb H}}
\nc{\LL}{{\Bbb L}}
\nc{\NN}{{\Bbb N}}
\nc{\PP}{{\Bbb P}}
\nc{\QQ}{{\Bbb Q}}
\nc{\RR}{{\Bbb R}}
\nc{\TT}{{\Bbb T}}
\nc{\VV}{{\Bbb V}}
\nc{\ZZ}{{\Bbb Z}}

\nc{\frakg}{{\frak g}}

\nc{\redtext}[1]{\textcolor{red}{#1}}
\nc{\bluetext}[1]{\textcolor{blue}{#1}}

\begin{document}

\title{Birkhoff type decompositions and the Baker--Campbell--Hausdorff recursion}

\author{Kurusch Ebrahimi-Fard}
\address{I.H.\'E.S.,
         Le Bois-Marie,
         35, Route de Chartres,
         F-91440 Bures-sur-Yvette, France}
         \email{kurusch@ihes.fr}
         \urladdr{http://www.th.physik.uni-bonn.de/th/People/fard/}

\author{Li Guo}
\address{Department of Mathematics and Computer Science,
         Rutgers University,
         Newark, NJ 07102, U.S.A.}
         \email{liguo@newark.rutgers.edu}
         \urladdr{http://newark.rutgers.edu/~liguo/}

\author{Dominique Manchon}
\address{Universit\'e Blaise Pascal,
         C.N.R.S.-UMR 6620,
         63177 Aubi\`ere, France}
         \email{manchon@math.univ-bpclermont.fr}
         \urladdr{http://math.univ-bpclermont.fr/~manchon/}

\date{March 14, 2006\\ \noindent {\footnotesize{${}\phantom{a}$ 2001 PACS Classification:
03.70.+k, 11.10.Gh, 02.10.Hh}} }

\begin{abstract}
We describe a unification of several apparently unrelated
factorizations arisen from quantum field theory, vertex operator
algebras, combinatorics and numerical methods in differential
equations. The unification is given by a Birkhoff type
decomposition that was obtained from the
Baker--Campbell--Hausdorff formula in our study of the Hopf
algebra approach of Connes and Kreimer to renormalization in
perturbative quantum field theory. There we showed that the
Birkhoff decomposition of Connes and Kreimer can be obtained from
a certain Baker--Campbell--Hausdorff recursion formula in the
presence of a Rota--Baxter operator. We will explain how the same
decomposition generalizes the factorization of formal exponentials
and uniformization for Lie algebras that arose in vertex operator
algebra and conformal field theory, and the even-odd decomposition
of combinatorial Hopf algebra characters as well as to the Lie
algebra polar decomposition as used in the context of the
approximation of matrix exponentials in ordinary differential
equations.
\end{abstract}

\maketitle

\tableofcontents


\section{Introduction}
\label{sect:intro}

The results presented in this paper grew out of an extension of the
study on Rota--Baxter algebras and their applications to areas of
mathematics and physics, including quantum field theory, classical
integrable systems, number theory, operads, combinatorics and Hopf
algebras.

In recent works of Connes and
Kreimer~\cite{CK1,CK2,CK3,KreimerChen}, triggered by Kreimer's
seminal paper~\cite{KreimerHopf}, new progresses were made in the
understanding of the process of renormalization in perturbative
quantum field theory, both in terms of its mathematical and its
physical contents. These results motivated further studies, among
other directions, in the context of Rota--Baxter
algebras~\cite{E-G-K2,E-G-K3,EG,EGGV}. We refer the reader to
\cite{EK,FG,KreimerRev,Ma} for more details and references in this
field.

One key result in the works of Connes and Kreimer is the Birkhoff
decomposition of Feynman rules that captures the process of
renormalization. Working in a fully algebraic framework of
complete filtered Rota--Baxter algebras, it was shown
in~\cite{E-G-K2,E-G-K3} that the Connes--Kreimer decomposition
follows from an additive decomposition in a Rota--Baxter (Lie)
algebra through the exponential map. Thereby the well-known
Bogoliubov formulae, which form the backbone of the standard BPHZ
renormalization procedure \cite{BoPa57,Hepp66,Zimmermann69}, were
derived as a special case of a generalization of Spitzer's
classical identity. As a side remark we mention here that a
similar factorization was independently established as a
fundamental theorem for Lie algebras in integrable systems
\cite{BBT,STS83,STS00}.

These results rely in part on general properties of Rota--Baxter
operators, but also on a recursive equation based on the famous
Baker--Campbell--Hausdorff formula. We will show that in certain
favorable cases we are able to give the recursion in closed form.
The main topic of this paper is the exploration of further
applications of this recursive equation, which was dubbed the
$BC\!H$-recursion. We will show its appearance in several fields.
First, in the context of Rota--Baxter algebra, it is shown to be a
generalization of the Magnus expansion known from matrix initial
value problems. Then, by applying the recursion to the
decomposition for certain Lie algebras, we derive the
factorization of formal exponentials and uniformization in the
work of Barron, Huang and Lepowsky~\cite{B-H-L} which is itself a
generalization of several of their earlier results. Furthermore,
we link explicitly the $BC\!H$-recursion with the even-odd
decomposition of characters of connected graded Hopf algebras
derived in recent work of Aguiar, Bergeron and
Sottile~\cite{A-B-S}. This way we achieve an exponential form of
their decomposition. This result also relates to our last point.
We give a simplified approach to the polar decomposition in the
work of Munthe-Kaas and collaborators~\cite{Za1,Za2,Za3} on
numerical solutions of differential equations.\\

Let us outline the organization of this paper. After the above
introduction, Section~\ref{sect:SetUp} provides the key result. In
subsection~\ref{subsect:assoAlg} we introduce a certain
Baker--Campbell--Hausdorff type equation as the main object of
this work, together with a general factorization theorem for
complete filtered associative and Lie algebras. Solutions to this
recursion are given under particular assumptions.
Subsection~\ref{subsect:Rota-Baxter} combines the former results
with the notion of Rota--Baxter algebra, giving rise to a
generalization of Spitzer's classical identity. We relate the
$BC\!H$-recursion to Magnus' expansion in the context of weight
zero Rota--Baxter maps. As a motivational example we recall in
Section~\ref{sect:CKqft} how this, together with Atkinson's
factorization theorem, applies to the work of Connes and Kreimer
in perturbative quantum field theory. Section~\ref{sect:BHL}
relates our findings to the work of Barron, Huang and Lepowsky on
the factorization of formal exponentials and uniformization. After
that, we deduce in Section~\ref{sect:ABS} the even-odd
decomposition of combinatorial Hopf algebra characters defined by
Aguiar, Bergeron and Sottile and give a closed form for the
$BC\!H$-recursion in this particular setting. Finally, in
Section~\ref{sect:Zanna} using similar ideas we briefly mention a
simplified approach to some results in the work of Munthe-Kaas and
collaborators.


\section{The general set up}
\label{sect:SetUp}

In the following $\mathbb{K}$ denotes the base field of
characteristic zero, over which all algebraic structures are
defined. Many results remain true if it is replaced by a
commutative $\mathbb{Q}$-algebra.\smallskip

Here we establish general results to be applied in later sections.
We start with a complete filtered associative algebra $A$ together
with a filtration preserving linear map $P$ on $A$ as general
setting. We obtain from the Baker--Campbell--Hausdorff ($\BCH$)
series a non-linear map $\chi$ on $A$ which we called
$BC\!H$-recursion in~\cite{E-G-K2,E-G-K3,EG}. This recursion gives
a decomposition on the exponential level, and a one-sided inverse
of the $\BCH$ series with the later regarded as a map from $A
\times A \to A$. The results naturally apply in the Lie algebra
case. We then consider the above setting in the realm of a
complete filtered associative Rota--Baxter algebra giving rise to
a generalization of Spitzer's classical identity. We conclude this
section by showing how the $BC\!H$-recursion can be seen as a
generalization of the Magnus expansion in the context of
Rota--Baxter algebras.


\subsection{The Baker--Campbell--Hausdorff recursion}
\label{subsect:assoAlg}

Let $A$ be a complete filtered associative algebra. Thus $A$ has a
decreasing filtration $\{A_n\}$ of sub-algebras such that $A_mA_n
\subseteq A_{m+n}$ and $A \cong \invlim A/A_n$ (i.e., $A$ is
complete with respect to the topology from $\{A_n\}$). For
instance, consider for $A$ being an arbitrary associative algebra,
the power series ring $\mathcal{A}:=A[[t]]$ in one (commuting)
variable $t$. Another example is given by the subalgebra
$\mathcal{M}^\ell_n(A) \subset \mathcal{M}_n(A)$ of strictly
(upper) lower triangular matrices in the algebra of $n \times n$
matrices with entries in $A$, and with $n$ finite or infinite. By
the completeness of $A$, the functions
$$
    \exp: A_1 \to 1+A_1,\qquad \exp(a)=\sum_{n=0}^\infty \frac{a^n}{n!},
$$
$$
    \log: 1+A_1 \to A_1,\qquad \log (1+a)=-\sum_{n=1}^\infty \frac{(-a)^n}{n}
$$
are well-defined and are the inverse of each other.

The Baker--Campbell--Hausdorff formula is the power series
$\BCH(x,y)$ in the non-commutative power series algebra
$A:=\QQ\langle\langle x,y \rangle\rangle$ (which is the free
noncommutative complete $\QQ$-algebra with generators $x, y$) such
that~\cite{Reutenauer,Varadarajan}
$$
    \exp(x)\exp(y)=\exp\big(x+y+\BCH(x,y)\big).
$$
Let us recall the first few terms of $\BCH(x,y)$ which are
$$
    \BCH(x,y)=\frac{1}{2}[x,y] + \frac{1}{12}[x,[x,y]] - \frac{1}{12}[y,[x,y]] -
                                                     \frac{1}{24}[x,[y,[x,y]]] +\cdots
$$
where $[x,y]:=xy-yx$ is the commutator of $x$ and $y$ in $A$. Also
denote $C(x,y):=x+y+\BCH(x,y)$. So we have
$$
    C(x,y) = \log \big(\exp(x)\exp(y)\big),
$$
which is a special case of the Hausdorff series \cite{Loday94}
$$
    Z(x_1,\dots,x_n):=\log \big(\exp(x_1)\cdots\exp(x_n)\big).
$$
Then for any complete $\QQ$-algebra $A$ and $u,v \in A_1$, $C(u,v)
\in A_1$ is well-defined. So we get a map
$$
    C : A_1 \times A_1 \to A_1.
$$

Now let $P: A \to A$ be any linear map preserving the filtration
of $A$. We define $\tilde{P}$ to be $\id_A-P$. For $a \in A_1$,
define $\chi(a) = \lim_{n \to \infty} \chi_{(n)}(a)$ where
$\chi_{(n)}(a)$ is given by the $BC\!H$-recursion
\allowdisplaybreaks{
\begin{eqnarray}
      \chi_{(0)}(a) &:=& a, \nonumber\\
    \chi_{(n+1)}(a) & =& a - \BCH \big( P(\chi_{(n)}(a)),(\id_A-P)(\chi_{(n)}(a)) \big),
    \label{eq:chik}
\end{eqnarray}}
and where the limit is taken with respect to the topology given by
the filtration. Then the map $\chi: A_1 \to A_1$ satisfies
\begin{equation}
    \label{BCHrecursion1}
    \chi(a) = a - \BCH\big(P(\chi(a)),\tilde{P}(\chi(a))\big).
\end{equation}
This map appeared in \cite{E-G-K2,E-G-K3,EG}, where also more
details can be found. The following proposition gives further
properties of the map $\chi$, improving a result in~\cite{Ma} (in
the {\texttt{arXiv}} version, Paragraph II.6.).

\begin{prop} \label{prop:conv}
For any linear map $P: A \to A$ preserving the filtration of $A$
there exists a unique (usually non-linear) map $\chi: A_1 \to A_1$
such that $(\chi - \id_A)(A_i) \subset A_{2i}$ for any $i \ge 1$,
and such that, with $\tilde{P}:= \id_A - P$ we have
\begin{equation}
    \label{truc}
    \forall a \in A_1,\quad a = C \Big(P\big(\chi(a)\big),\,\tilde{P}\big(\chi(a)\big)\Big).
\end{equation}
This map is bijective, and its inverse is given by
\begin{equation}
    \label{chi-inverse}
    \chi^{-1}(a)=C\big(P(a),\,\tilde{P}(a)\big)
                        =a+\BCH\big(P(a),\,\tilde{P}(a)\big).
\end{equation}
\end{prop}

\begin{proof}
Equation (\ref{truc}) can be rewritten as
$$
    \chi(a)=F_a \big(\chi(a)\big),
$$
with $F_a: A_1 \to A_1$ defined by
$$
    F_a(b) = a - \BCH\big(P(b),\tilde{P}(b)\big).
$$
This map $F_a$ is a contraction with respect to the metric
associated with the filtration: indeed if $ b, \varepsilon \in A$
with $\varepsilon \in A_n$, we have
$$
    F_a(b + \varepsilon) - F_a(b) = \BCH\big(P(b),\,\tilde{P}(b)\big)
                                            -\BCH\big(P(b+\varepsilon),\,\tilde{P}(b+\varepsilon)\big).
$$
The right-hand side is a sum of iterated commutators in each of
which $\varepsilon$ does appear at least once. So it belongs to
$A_{n+1}$. So the sequence $F_a^n(b)$ converges in $A_1$ to a
unique fixed point $\chi(a)$ for $F_a$.

Let us remark that for any $a \in A_i$, then, by a straightforward
induction argument, $\chi_{(n)}(a)\in A_i$ for any $n$, so
$\chi(a)\in A_i$ by taking the limit. Then
$\chi(a)-a=\BCH\Big(P\big(\chi(a)\big),\, \tilde
P\big(\chi(a)\big)\Big)$ clearly belongs to $A_{2i}$. Now consider
the map $\psi:A_1\to A_1$ defined by
$\psi(a)=C\big(P(a),\,\tilde{P}(a)\big)$. It is clear from the
definition of $\chi$ that $\psi \circ \chi = \hbox{id}_A$. Then
$\chi$ is injective and $\psi$ is surjective. The injectivity of
$\psi$ will be an immediate consequence of the following lemma

\begin{lem}\label{lem:dilation}
The map $\psi$ increases the ultrametric distance given by the
filtration.
\end{lem}

\begin{proof}
For any $x,y \in A_1$ the distance $d(x,y)$ is given by
${\rm{e}}^{-n}$ where $n=\hbox{sup}\{k\in\NN,\,x-y\in A_k\}$. We
have then to prove that $\psi(x)-\psi(y)\notin A_{n+1}$. But
\allowdisplaybreaks{
\begin{eqnarray*}
  \psi(x)-\psi(y) &=& x-y+\BCH\big(P(x),\,\tilde P(x)\big)-\BCH\big(P(y),\,\tilde{P}(y)\big)\\
                  &=& x-y+\Big(\BCH\big(P(x),\,\tilde{P}(x)\big)
                        - \BCH\big(P(x)-P(x-y),\,\tilde{P}(x) - \tilde{P}(x-y)\big)\Big).\\
\end{eqnarray*}}
The rightmost term inside the large brackets clearly belongs to
$A_{n+1}$. As $x-y\notin A_{n+1}$ by hypothesis, this proves the
claim.
\end{proof}
The map $\psi$ is then a bijection, so $\chi$ is also bijective,
which proves Proposition \ref{prop:conv}.
\end{proof}

Now let $\frakg$ be a complete filtered Lie algebra. Let
$A:=\mathcal{U}(\frakg)$ be the universal enveloping associative
algebra of $\frakg$. Then with the induced filtration from
$\frakg$, $A$ is a complete filtered associative algebra. $A$ is
also a complete Lie algebra under the bracket $[x,y]:=xy-yx$ and
contains $\frakg$ as a complete filtered sub-Lie algebra. Let $P:
\frakg \to \frakg$ be a linear map preserving the filtration in
$\frakg$. We can extend $P$ to a linear map $\hat{P}: A \to A$
that preserves the filtration in $A$. A simple way to build such
an extension (by no means unique) is to choose any supplementary
subspace $V$ of $\frak{g}$ inside $A$ and to extend $P$ by the
identity map on the complement. If $P$ is idempotent, so is
$\hat{P}$. As is well-known~\cite{Reutenauer,Varadarajan}, the
power series $C(x,y)$ and $\BCH(x,y) \in \QQ\langle\langle
x,y\rangle\rangle$ are Lie series. Therefore, the map $\chi: A_1
\to A_1$ in Eq.~(\ref{BCHrecursion1}) and
Proposition~\ref{prop:conv} restricts to a bijective map
$$
    \chi: \frakg_1 \to \frakg_1
$$
with its inverse given by Eq.~(\ref{chi-inverse}). Further, for
$u,v \in \frak g_1$, $C(u,v)$ is a well-defined element in
$\frakg_1$. We thus have
$$
    C:\frakg_1\times \frakg_1 \to \frakg_1
$$
as in the associative case.

The following theorem contains the key result of our exposition.
It states a general decomposition on $A$ implied by the map
$\chi$.

\begin{thm} \label{thm:bch}
Let $A$ be a complete filtered associative algebra or Lie algebra
with a linear, filtration preserving map $P: A \to A$.
\begin{enumerate}
    \item \label{it1:bch-thm} For any $a \in A_1$, we have
        \begin{equation}
            \exp(a)=\exp\big(P(\chi(a))\big)\exp\big(\tilde{P}(\chi(a))\big).
            \label{eq:bch}
        \end{equation}

    \item\label{it2:bch-thm} $C: A_1 \times A_1 \to A_1$ has a right inverse $D_P$ given by
        $$
            D_P=(P \circ \chi,\ \tilde{P} \circ \chi): A_1 \to A_1 \times A_1.
        $$

    \item \label{it3:bch-thm} $C$ restricts to a bijection
        $$
            C: D_P(A_1) \to A_1.
        $$

    \item \label{it4:bch-thm} Furthermore, for any subset $B$ of $A_1$, $C$ restricts to a bijection
        $$
            C: D_P(B) \to B.
        $$
\end{enumerate}
\end{thm}

\begin{proof}
(\ref{it1:bch-thm}) follows since
$$
    C\big(P(\chi(a)),\ \tilde{P}(\chi(a))\big)=a.
$$
(\ref{it2:bch-thm}) follows since
$$
    C\circ D_P(a) = C\big(P(\chi(a)),\ \tilde{P}(\chi(a))\big)=a.
$$
(\ref{it3:bch-thm}) is a general property of maps:
$$
    (D_P \circ C)\Big|_{D_P(A_1)}(a_1,a_2)=(D_P\circ C)\circ D_P(a)
                                          = D_P\circ (C\circ D_P)(a)
                                          = D_P(a).
$$
(\ref{it4:bch-thm}) is clear as $D_P$ is a (two-sided) inverse for
the restriction of $C$ to $D_P(A_1)$.
\end{proof}

The particular case when the map $P$ is idempotent deserves
special attention.

\begin{thm}\label{thm:idemp}
Let $P: A \to A$ be an idempotent linear map preserving the
filtration of $A$. Let $A = A_- \oplus A_+$ be the corresponding
vector space decomposition, with $A_- := P(A)$ and $A_+ :=
\tilde{P}(A)$. Let $A_{1,-}:=P(A_1)$ and
$A_{1,+}:=\tilde{P}(A_1)$. Let $\chi: A_1 \to A_1$ be the
$BC\!H$-recursion map associated to the map $P$, and let
$\tilde{\chi}: A_1 \to A_1$ be the $BC\!H$-recursion map
associated to $\tilde{P}$
\begin{enumerate}
    \item \label{it:factor} {\rm{(Factorization Theorem)}} $C$ restricts to a
          bijection
            $$
                C_-: A_{1,-} \times A_{1,+} \longrightarrow A_1.
            $$

    \item \label{it:uniform} {\rm{(Formal Uniformization Theorem)}} There
           exists a unique bijection
            $$
                \Psi: A_{1,+} \times A_{1,-} \longrightarrow A_{1,-} \times A_{1,+}
            $$
            such that for $a = (a_+,a_-)\in A_{1,+} \times A_{1,-}$, we have
            $$
                \exp(a_+)\exp(a_-) =\exp(\pi_-(\Psi(a)))\exp(\pi_+(\Psi(a))),
            $$
            where $\pi_{\pm}: A_{1,-}\times A_{1,+}\to A_{1,\pm}$ are the
            projectors.
    \item \label{it:closed} The inverse map of $C_-$ in part (\ref{it:factor}) is given by
            $$
                D_P(a)=\Big(P\big(\chi(a)\big),\,\tilde P\big(\chi(a)\big)\Big),
                                                                    \qquad a\in A_1,
            $$
            and the uniformization map $\Psi$ in part (\ref{it:uniform}) writes
            $$
                \Psi(a)=\Big(P\big(\chi\circ C(a)\big),\,\tilde P\big(\chi\circ C(a)\big)\Big)
            $$
            or
            $$
                \Psi(a)=\Big(P\big(\chi\circ \tilde \chi^{-1}(a_++a_-)\big),\,
                                    \tilde P\big(\chi \circ \tilde\chi^{-1}(a_++a_-)\big)\Big)
            $$
            with $a=(a_+,a_-) \in A_{1,+}\times A_{1,-}$.
\end{enumerate}
\end{thm}

The statements (\ref{it:factor}) and (\ref{it:uniform}) in the
above theorem generalize theorems of Barron, Huang and
Lepowsky~\cite{B-H-L} which are themselves generalizations of
factorization and uniformization theorems for Lie algebras and Lie
superalgebras such as Virasoro algebras and Neveu-Schwarz
algebras, respectively. This is the motivation for the naming of
those items. See Section~\ref{sect:BHL} for further details.

\begin{proof}
(\ref{it:factor}) We already know from item (\ref{it3:bch-thm}) of
Theorem \ref{thm:bch} that $D_P$ is a right inverse for $C_-$. But
it is also a left inverse, as for any $(x,y) \in A_{1,-} \times
A_{1,+}$ there is a unique $v \in A_1$ such that $x = P(v)$ and $y
= \tilde{P}(v)$, and we have \allowdisplaybreaks{
\begin{eqnarray*}
    D_P \circ C(x,y) &=& D_P \circ C\big(P(v),\,\tilde{P}(v)\big)   \\
                   &=& D_P \circ \chi^{-1}(v)                       \\
                   &=& (P,\ \tilde{P})(v)                             \\
                   &=& (x,y).
\end{eqnarray*}}

\noindent (\ref{it:uniform}) By the same argument as for part
(\ref{it:factor}), $C$ restricts to a bijection
$$
    C_+: A_{1,+}\times A_{1,-} \longrightarrow A_1.
$$
Its inverse is now $D_{\tilde{P}}$. Since $A_1=A_{1,+}\oplus
A_{1,-}=A_{1,-}\oplus A_{1,+}$, we can define $\Psi$ by the
following diagram
$$
    \xymatrix{ A_{1,+}\times A_{1,-} \ar^{C_+}[rr]
                \ar^{\Psi}[d] && A_{1,+}\oplus A_{1,-} \ar_{\sigma}[d]\\
                   A_{1,-} \times A_{1,+} \ar^{C_-}[rr]&&
                   A_{1,-} \oplus A_{1,+}
            }
$$
Here $\sigma$ is just a cosmetical way to write the identity map
$$
    \sigma(b^+ + b^-)=b^-+ b^+.
$$
$\Psi$ is bijective since $C_+$ and $C_-$ in the diagram are. We
see also from the diagram that this $\Psi$ is the unique map such
that
$$
    \exp(a_+)\exp(a_-) =\exp\big(\pi_-(\Psi(a))\big)\exp\big(\pi_+(\Psi(a))\big).
$$
(\ref{it:closed}) This follows from the above commutative diagram
and part (\ref{it:factor}): now we can compute
\allowdisplaybreaks{
\begin{eqnarray*}
    \Psi(a)&=& D_P\circ C_+(a)\\
           &=& \big(P\circ\chi\circ C_+(a),\,P\circ\chi\circ C_+(a)\big)\\
           &=& \big(P(\chi\circ\tilde\chi^{-1}(a_++a_-)),\,\tilde{P}(\chi\circ\tilde\chi^{-1}(a_++a_-))\big)
\end{eqnarray*}}
which ends the proof of Theorem \ref{thm:idemp}.
\end{proof}

\begin{cor}\label{decomposition}
Under the hypotheses of Theorem \ref{thm:idemp}, for any $\eta \in
1 + A_1$ there are unique $\eta_{-} \in \exp\big(A_{1,-}\big)$ and
$\eta_{+} \in \exp\big(A_{1,+}\big)$ such that $\eta = \eta_-\,
\eta_+$.
\end{cor}

\begin{proof}
This follows directly from the first item of Theorem~\ref{thm:bch}
and the first item of Theorem~\ref{thm:idemp}, as the exponential
map is a bijection from $A_1$ onto $1+A_1$.
\end{proof}

Let us finish this section with two observations simplifying the
$BC\!H$-recursion considerably. The first one is of more general
character. To begin with it might be helpful to work out the first
few terms of the recursion for the map $\chi$ in (\ref{eq:chik}).
For this let us introduce a dummy parameter $t$ and write
$\chi(at)= t \sum_{k \geq 0} \chi^{(k)}(a)t^k$. For $k=0,1,2$ we
readily find $\chi^{(0)}(a)=a$ and \allowdisplaybreaks{
\begin{eqnarray}
    \chi^{(1)}(a) &=& - \frac{1}{2}[P(a),\tilde{P}(a)]=
                                    - \frac{1}{2}[P(a),a] \label{chi1}\\
    \chi^{(2)}(a) &=& -\frac{1}{2}[P(\chi^{(1)}(a)),\tilde{P}(a)] - \frac{1}{2}[P(a),\tilde{P}(\chi^{(1)}(a))]  \nonumber\\
         & & \hspace{4.5cm} - \frac{1}{12}\Big( \big[P(a),[P(a),a]\big] - \big[\tilde{P}(a),[P(a),a]\big] \Big) \nonumber\\
         &=& +\frac{1}{4}\big[P([P(a),a]),\tilde{P}(a)\big] + \frac{1}{4}\big[P(a),\tilde{P}([P(a),a])\big]     \nonumber\\
         & & \hspace{4.5cm}  -\frac{1}{12}\Big( \big[P(a),[P(a),a]\big] - \big[\tilde{P}(a),[P(a),a]\big] \Big) \nonumber\\
         &=& \frac{1}{4}\big[P([P(a),a]),a\big] + \frac{1}{12}\Big( \big[P(a),[P(a),a]\big]
                                                                                - \big[[P(a),a],a\big] \Big).\label{chi2}
\end{eqnarray}}

\noindent In both the last cases $\tilde{P}=\id_A-P$ has
completely disappeared. Therefore, we might expect to find a
simpler recursion for the map $\chi$, without the appearance of
$\tilde{P}$. Indeed, such a simplification follows using the
factorization property, implied by the $\chi$ map on $A$ in item
(\ref{it1:bch-thm}) of Theorem~\ref{thm:bch}.

\begin{lem}\cite{E-G-K3}
\label{simpleCHI} Let $A$ be a complete filtered algebra and $P: A
\to A$ a linear map preserving the filtration. The map $\chi$ in
(\ref{BCHrecursion1}) solves the following recursion
 \begin{equation}
   \label{BCHrecursion2}
   \chi(u):=u + \BCH\big( -P(\chi(u)),u \big),\;\; u \in A_1.
 \end{equation}
\end{lem}

\begin{proof}
For any element $u \in A$  we can write $u = P(u) +
(\id_{A}-P)(u)$ using linearity of $P$. The definition of the map
$\chi$ then implies for $u \in A_1$ that $\exp(u)=
\exp\big(P(\chi(u))\big) \exp\big(\tilde{P}(\chi(u))\big)$, see
Eq.~(\ref{eq:bch}). Furthermore, \allowdisplaybreaks{
\begin{eqnarray*}
  \exp\big(\tilde{P}(\chi(u))\big)  &=&  \exp\big(-P(\chi(u))\big) \exp(u) \\
                                    &=& \exp\big(-P(\chi(u)) + u + \BCH(-P(\chi(u)),u)\big).
\end{eqnarray*}}
Bijectivity of the $\exp$ map then implies that
 \allowdisplaybreaks{
\begin{eqnarray*}
 \chi(u) - P(\chi(u)) &=& -P(\chi(u)) + u + \BCH\big(-P(\chi(u)),u\big).
\end{eqnarray*}}
From which Equation (\ref{BCHrecursion2}) follows.
\end{proof}

Our second observation is of more particular type. Again, it
concerns the linear map $P$ in the definition of the
$BC\!H$-recursion $\chi$. We will treat a special case, providing
a solution, i.e., closed form, for the $BC\!H$-recursion. Further
below in Section~\ref{sect:ABS} we will observe another instance
where a closed form for the $BC\!H$-recursion can be derived, see
Eq.~(\ref{chiSimple2}).

Let us now assume that the linear map $P: A \to A$ in the
$BC\!H$-recursion in Eq.~(\ref{BCHrecursion2}) of Lemma
\ref{simpleCHI} is an idempotent map, and moreover that it
respects multiplication in $A$. This makes $P$ respectively
$\tilde{P}=\id_A - P$ a Rota--Baxter map to be introduced in the
following section although the map $\tilde{P}$ is not an algebra
morphism. We then have

\begin{lem} \label{lem:algIdemRB}
Let $A$ be a complete filtered associative algebra with filtration
preserving linear map $P: A \to A$, which moreover is an
idempotent algebra homomorphism. Then the map $\chi$ in
Eq.~(\ref{BCHrecursion2}) of Lemma \ref{simpleCHI} has the simple
form
 \begin{equation}
   \label{BCHrecur3}
   \chi(u) = u + \BCH\big( -P(u),u \big),
 \end{equation}
for any element $u \in A_1$.
\end{lem}

\begin{proof}
The proof follows from Lemma~\ref{simpleCHI}, since
$P(\chi(u))=P(u)$. The latter results from the multiplicativity of
$P$, i.e., applying $P$ to Eq.~(\ref{BCHrecursion1}) we obtain
$$
    P(\chi(u))=P(u) + \BCH\big(P^2(\chi(u)),(P\circ\tilde{P})(u))\big).
$$
Since $P$ is idempotent, we have $P\circ \tilde{P}=P-P^2=0$. Thus
$P(\chi(u))=P(u)$.
\end{proof}

\begin{remark} With the foregoing assumptions on $P$ the factorization in item
(\ref{it1:bch-thm}) of Theorem \ref{thm:bch} simplifies
considerably. For any $a \in A_1$, we have
\begin{equation}
    \exp(a)=\exp\big(P(a)\big) \exp\big(\tilde{P}(a) + \BCH\big( -P(a),a \big)\big).
    \label{eq:simpleFact}
\end{equation}
\end{remark}


\subsection{Rota--Baxter operator}
\label{subsect:Rota-Baxter}

In the 1950s and early 1960s, several interesting results were
obtained in the fluctuation theory of probability. One of the most
well-known is Spitzer's identity~\cite{Spitzer}. In a seminal 1960
paper~\cite{Baxter}, the American mathematician G. Baxter deduced
it from a certain operator identity, that later bore his name.
During the early 1960s and 1970s, algebraic, combinatorial and
analytic aspects of Baxter's work were studied by several people,
among them G.-C.~Rota and F.~V.~Atkinson. Much of the recent
renewed interest into these works owes to Rota's later survey
articles~\cite{Rota2,Rota3} and talks during the 1990s. Related
concepts were independently developed by Russian physicists during
the 1980s. Especially in Belavin and Drinfeld's, and
Semenov-Tian-Shansky's papers~\cite{BelavinDrinfeld1,STS83} on
solutions of the (modified) classical Yang--Baxter equation. In
this context let us mention another connection linked with the
last remark. Aguiar~\cite{Aguiar} related Rota--Baxter operators
of weight zero to the associative analog of the classical
Yang--Baxter equation, which also appeared in~\cite{Polish}.
\smallskip

Now we assume that $A$ is an associative algebra and $P$ a
Rota--Baxter operator of weight $\theta$ satisfying the
Rota--Baxter relation
\begin{equation}
        P(x)P(y) + \theta P(xy) = P\big(xP(y)\big) + P\big(P(x)y\big)
        \label{eq:RB}
\end{equation}
for all $x,y \in A$ \cite{Baxter,Rota1,Rota2,RotaSmith}. A
Rota--Baxter algebra of weight $\theta$ is an algebra with a
Rota--Baxter operator denoted by the pair $(A,P)$. The operator
$\tilde{P} := \theta \: \id_A - P$ also is a Rota--Baxter map of
weight $\theta$, such that the mixed relation
\begin{equation}
      \label{eq:mixedRB}
      P(x)\tilde{P}(y)=\tilde{P}\big(P(x)y\big) + P\big(x\tilde{P}(y)\big)
\end{equation}
is satisfied for all $x,y \in A$. The image of $P$ as well as
$\tilde{P}$ are subalgebras in $A$. A Rota--Baxter
ideal $I$ is an ideal $I$ of $A$ such that $P(I)
\subseteq I$.

The case $\theta = 0$ corresponds to the integration by parts
property of the usual Riemann integral. An important class of
examples is given by idempotent Rota--Baxter maps, i.e.,
projectors, where identity (\ref{eq:RB}) (of weight $\theta=1$)
implies that the Rota--Baxter algebra $A$ splits as a direct sum
into two parallel subalgebras given by the image, respectively
kernel, of $P$. Assuming $P$ to be an idempotent algebra morphism
is sufficient to imply that it is a Rota--Baxter map. As an
example of an idempotent Rota--Baxter map which is moreover an
algebra morphism, truncate the Taylor expansion of a real function
at a point $a$ at zeroth order, i.e., evaluate a real function at
a point $a$, $P_a^{(0)}(f)(x)=f(a)$.

The modified Rota--Baxter operator, $B := \theta\:\id_A - 2P$,
satisfies the modified Rota--Baxter relation
\begin{equation}
  B(x)B(y) + \theta^2 xy = B\big(B(x)y + xB(y)\big).
\end{equation}
for all $x$ and $y$ in $A$. For a modified Rota--Baxter operator
$B$ coming from an idempotent Rota--Baxter map $P$, we have $B^2 =
\id_A$, $B\circ P= -P$, and $B\circ \tilde{P}= \tilde{P}$.

Taking the Lie algebra associated to $(A,P)$, with commutator
bracket $[x,y]:= xy-yx$, for all $x,y \in A$, we find the
Rota--Baxter Lie algebra, $(L_A,P)$, of weight $\theta$ with $P$
fulfilling
\begin{equation}
   \label{eq:RBLie}
   [P(x),P(y)] + \theta P([x,y]) = P\big([P(x),y] + [x,P(y)]\big).
\end{equation}
Similarly, for the modified Rota--Baxter map. Both equations are
known as (the operator form of) the (modified) classical
Yang--Baxter\footnote{Refereing to the Australian physicist Rodney
Baxter.} equations~\cite{BelavinDrinfeld1,STS83}.

Every Rota--Baxter algebra $(A,P)$ of weight $\theta$ allows for a
new product defined in terms of the Rota--Baxter map $P$
\begin{equation}
   a *_P b :=P(a)b + a P(b) - \theta ab
   \label{doubleRB}
\end{equation}
such that the vector space $A$ with this product is a Rota--Baxter
algebra of the same weight, with $P$ as its Rota--Baxter map. We
will denote it by $(A_P,P)$. The Rota--Baxter map $P$ becomes an
(not necessarily unital) algebra homomorphism from $A_P$ to $A$,
$P(a *_P b) = P(a)P(b)$. For $\tilde{P}$ we have $\tilde{P}(a *_P
b)= -\tilde{P}(a) \tilde{P}(b)$.

A complete filtered Rota--Baxter algebra is defined to be a
Rota--Baxter algebra $(A,P)$ with a complete filtration by
Rota--Baxter ideals $\{A_n\}$. Again, consider for any weight
$\theta$ Rota--Baxter algebra $(A,P)$ the power series ring
$\mathcal{A}:=A[[t]]$ and define an operator $\mathcal{P}:
\mathcal{A} \to \mathcal{A}$, $\mathcal{P}(\sum_{n=0}^\infty a_n
t^n):=\sum_{n=0}^\infty P(a_n)t^n$. Then $(\mathcal{A},
\mathcal{P})$ is a complete filtered Rota--Baxter algebra of
weight $\theta$. In the case of the algebra of strictly (upper)
lower triangular matrices $\mathcal{M}^\ell_n(A)$ with $n\leq
\infty$ and entries in a weight $\theta$ Rota--Baxter algebra
$(A,P)$, define the Rota--Baxter map $\mathcal{P}:
\mathcal{M}^\ell_n(A) \to \mathcal{M}^\ell_n(A)$ entrywise,
$\mathcal{P}(\alpha) =  \big(P(\alpha_{ij})\big)$, for $\alpha$ in
$\mathcal{M}^\ell_n(A)$~\cite{EG}.

The normalized map $\theta^{-1} P$ is a Rota--Baxter operator of
weight one. In the following we will assume that any Rota--Baxter
map is of weight one, if not stated otherwise. The next
proposition contains the generalization of Spitzer's identity to
non-commutative Rota--Baxter algebras.

\begin{prop} \label{pp:sp} \cite{E-G-K2,E-G-K3,EG}
Let $(A,P)$ be a complete filtered Rota--Baxter algebra. The
factors on the right hand side of Equation (\ref{eq:bch})
\begin{equation}
    \exp(a) = \exp\big(P(\chi(a))\big) \exp\big(\tilde{P}(\chi(a))\big)
    \label{eq:exp}
\end{equation}
for $a \in A_1$ are the unique solutions to the equations
\begin{equation}
    u = 1 - P(\check{b}\ u) \quad {\rm resp.\ }\ u'=1 - \tilde{P}(u'\ \check{b}),
    \label{eq:inverseRecurs1}
\end{equation}
where $\check{b}:=\exp(-a)-1$ in $A_1$. Its inverses satisfy
uniquely the equations
\begin{equation}
    x = 1 - P(x\ b) \quad {\rm resp.\ }\ x'=1 - \tilde{P}(b\ x'),
    \label{eq:recurs}
\end{equation}
where $b:=\exp(a)-1=(1+\check{b})^{-1}-1 \in A_1$.
\end{prop}

The following theorem is due to Atkinson~\cite{Atkinson}.

\begin{thm} \label{thm:Atkinson} For the solutions $x$ and $x'$ in (\ref{eq:recurs})
(resp. its inverses in (\ref{eq:inverseRecurs1})) with
$b:=\exp(a)-1$ we have
\begin{equation}
     x (1+b) x' = 1, {\rm\ that\ is,\ } (1+b)=x^{-1}x'{}^{-1}.
    \label{eq:atkinson}
\end{equation}
If $P$ is idempotent, i.e., the algebra $A$ decomposes directly
into the images of $P$ and $\tilde{P}$, the factorization of $1+b$
is unique.
\end{thm}

The next corollary follows readily and is stated for completeness.

\begin{cor} \label{cor:bogoliubov}
Let $(A,P)$ be a complete filtered Rota--Baxter algebra. For the
solutions $u$ and $u'$ in (\ref{eq:inverseRecurs1}), we find the
equations
\begin{equation}
  \label{eq:inverseRecurs2}
  u = 1 + P\big(b\ x'), \qquad u' = 1 + \tilde{P}\big(x\ b),
\end{equation}
where $x$ and $x'$ are solutions of Eqs.~(\ref{eq:recurs}),
respectively.
\end{cor}

As a proposition we mention without giving further details the
fact that, using the double Rota--Baxter product $*_{P}$ in
(\ref{doubleRB}) for $\theta=1$, we may write
$$
  x = 1 + P\Big( \exp^{*_{P}}\big(-\chi(a) \big) - 1 \Big),
$$
where $\exp^{*_{P}}$ denotes the exponential defined in terms of
the product in (\ref{doubleRB}). This implies $-x\
b=\exp^{*_{P}}\big(-\chi(\log(1+b))\big) - 1$ for $1+b:=\exp(a)$.

When $(A,P)$ is commutative, the map $\chi$ reduces to the
identity map, giving back Spitzer's classical identity, for fixed
$b \in A_1$ \cite{Spitzer} \allowdisplaybreaks{
\begin{eqnarray}
  \exp\Big(- P\big(\log(1 + b)\big) \Big)
         &=&\sum_{n=0}^\infty  (-1)^n \underbrace{P\big(P( \cdots P(P}_{n-times}(b)b)\dots b)b\big),
        \label{SpitzerId}
\end{eqnarray}}
corresponding to the first recursion, $x = 1 - P(x \: b)$, in
(\ref{eq:recurs}). Replacing the Rota--Baxter map $P$ by the
identity map, the above identity reduces to the geometric series
for the element $-b \in A_1$. Proofs of this identity in the
commutative case have been given by quite a few authors, including
the aforementioned Atkinson~\cite{Atkinson},
Cartier~\cite{Cartier}, Kingman and Wendel~\cite{Kingman,Wendel}
as well as Rota and Smith~\cite{RotaSmith}. In fact,
Rota~\cite{Rota1} showed that this identity is equivalent to the
classical Waring identity relating elementary symmetric functions
and power symmetric functions.\\

\begin{remark} Coming back to Lemma \ref{lem:algIdemRB}, respectively Eq.~(\ref{eq:simpleFact})
we see immediately that in the case of a non-commutative
Rota--Baxter algebra $(A,P)$ with idempotent and multiplicative
Rota--Baxter map $P$ and thence necessarily of weight one,
implying $P(\chi(a))=P(a)$, for all $a \in A_1$, we have the
surprising result that the exponential solution to the recursion
$x = 1 - P(x\:b)$ in (\ref{eq:recurs}) can be written as a
geometric series
\begin{equation}
 \label{geomSeries}
 \exp\Big(-P\big(\chi(\log(1+b))\big)\Big)= \exp\Big(-P\big(\log(1+b)\big)\Big)=\frac{1}{1+P(b)}.
\end{equation}
Observe that the $BC\!H$-recursion $\chi$ disappeared after the
first equality, since $P(\chi(a))=P(a)$.\\
\end{remark}

The normalization of the weight one Rota--Baxter map $P$ to
$\theta P$ gives a Rota--Baxter map of weight $\theta$. This
implies the following modification of Proposition \ref{pp:sp}.

\begin{prop} \label{pp:BCHtheta}
For a weight $\theta \neq 0$ Rota--Baxter operator $P$, the map
$\chi$ in factorization (\ref{eq:exp}) of Proposition \ref{pp:sp}
generalizes to
\begin{equation}
    \chi_{\theta}(a) = a -
    \frac{1}{\theta}{\BCH}\Big(P\big(\chi_{\theta}(a)\big),\tilde{P}\big(\chi_{\theta}(a)\big)\Big).
    \label{BCH-recursion3}
\end{equation}
Similarly the recursion in Eq.~(\ref{BCHrecursion2}) of Lemma
\ref{simpleCHI} transposes into
\begin{equation}
    \label{BCH-recursion4}
    \chi_{\theta}(a)=a + \frac{1}{\theta}\BCH\Big(-P\big(\chi_{\theta}(a)\big),\theta a\Big),\;\; a \in A_1.
\end{equation}
 Such that for all $a \in A_1$ we have the decomposition
\begin{equation}
    \exp(\theta a) = \exp\big(P(\chi_\theta (a))\big) \exp\big(\tilde{P}(\chi_\theta (a))\big).
    \label{eq:exptheta}
\end{equation}
The factors on the right hand side of Equation (\ref{eq:exptheta})
are inverses of the unique solutions of the equations
\begin{equation}
    x = 1 - P(x\:b) \quad {\rm resp.\ }\ x'=1 - \tilde{P}(b\: x'),
    \label{eqs:special}
\end{equation}
where $1 + \theta b:=\exp(\theta a)$ in $A$.
\end{prop}

From this proposition we arrive at

\begin{cor} \label{cor:NC-SpitzerId-theta}
Spitzer's identity for a complete filtered non-commutative Rota--Baxter
algebra $(A,P)$ of weight $\theta \neq 0$ is
\begin{eqnarray}
  \exp\Big(- P\Big(\chi_{\theta}\Big(\frac{\log(1 + \theta b)}{\theta}\Big)\Big) \Big)
         &=&\sum_{n=0}^\infty  (-1)^n \underbrace{P\big(b P(b P(b \cdots
         P}_{n-times}(b))\dots)\big),
  \label{id:NC-SpitzerId-theta}
\end{eqnarray}
for $b \in A_1$.
\end{cor}

\noindent We call $\chi_{\theta}$ the $BC\!H$-recursion of weight
$\theta \in \mathbb{K}$, or simply $\theta$-$BC\!H$-recursion. As
we will see in the next part, the particular appearance of the
weight $\theta$ in
Eqs.~(\ref{BCH-recursion3},\ref{BCH-recursion4}) reflects the fact
that in the case of weight $\theta=0$, hence $\tilde{P}=-P$,
Atkinson's factorization formula (\ref{eq:atkinson}) in Theorem
\ref{thm:Atkinson} collapses to
\begin{equation}
    \label{AtkinsonZero}
    xx'=\big(1 - P(x\:b)\big)\big(1 + P(b\:x')\big)=1
\end{equation}
for any $b \in A_1$, which is in accordance with
(\ref{eqs:special}) for $\theta \to 0$.

\begin{remark}{\rm{ It should be clear that the decomposition in Eq.~(\ref{eq:exptheta})
in the above proposition is true for any complete filtered algebra
$A$ with filtration preserving linear map $P$ and
$\tilde{P}_{\theta}:=\theta \id_A - P$. Hence,
Eqs.~(\ref{BCH-recursion3}--\ref{BCH-recursion4}) generalize
Theorem \ref{thm:bch}. The Rota--Baxter property only enters in
the last part with respect to the equations in
(\ref{eqs:special}), respectively Corollary
\ref{cor:NC-SpitzerId-theta}. }}
\end{remark}


\subsection{The case of vanishing weight and the Magnus recursion}

Regarding Eq.~(\ref{AtkinsonZero}) in connection with the
factorization in Eq.~(\ref{eq:exptheta}) for a weight $\theta \neq
0$ Rota--Baxter algebra, it is interesting to observe the limit of
$\theta$ going to zero in formula (\ref{BCH-recursion4}) for the
$\theta$-$BC\!H$-recursion. The terms in the $\BCH$ series on the
right hand side of (\ref{BCH-recursion4}) vanish except for those
which are linear with respect to the second variable. In general
we may write $C(a,b) = a + b + \BCH(a,b)$ as a sum
~\cite{Reutenauer}
$$
  C(a,b)=\sum_{n \geq 0} H_n(a,b),
$$
where $H_n(a,b)$ is the part of $C(a,b)$ which is homogenous of
degree $n$ with respect to $b$. Especially, $H_0(a,b)=a$. For
$n=1$ we have
$$
    H_1(a,b)=\frac{\hbox{ad}\,a}{1-{\rm{e}}^{-\hbox{\eightrm ad}\,a}}(b).
$$
(see e.g. \cite{Go}). Hence we get a non-linear map $\chi_0$
inductively defined on the pro-nilpotent Lie algebra $A_1$ by the
formula \allowdisplaybreaks{
\begin{eqnarray}
    \chi_0(a)&=&-\frac{\hbox{ad}P\big(\chi_0(a)\big)}{1-{\rm{e}}^{\hbox{\eightrm ad}P(\chi_0(a))}}(a)
                \label{BCH-recur5a}\\
             &=& \bigg(1 + \sum_{n>0} b_n \Big[\hbox{\rm ad}P\big(\chi_0(a)\big)\Big]^{n}\bigg)(a)
                \label{BCH-recur5b}
\end{eqnarray}}
where $P$ is now a weight zero Rota--Baxter operator. We call this
the weight zero $BC\!H$-recursion. The coefficients
$b_n:=\frac{B_n}{n!}$ where $B_n$ are the Bernoulli numbers. For
$n=1,2,3,4$ we find the numbers $b_1=-1/2$, $b_2=1/12$, $b_3=0$
and $b_4=-1/720$. The first three terms in (\ref{BCH-recur5b}) are
\begin{equation}
    \label{pre-Magnus}
    \chi_0(a)= a - \frac 12[P(a),\,a]+\Big(\frac 14\big[P\big([P(a),\,a]\big),\,a\big]
    +\frac{1}{12}\big[P(a),\,[P(a),\,a]\big]\Big)+\cdots
\end{equation}

As a particular example we assume $P$ to be the Riemann integral
operator defined by $P\{a\}(x):=\int_0^{x}a(y)dy$, which is a
Rota--Baxter map of weight zero, i.e., it satisfies the
integration by parts rule
$$
 P\{a_1\}(x) P\{a_2\}(x) = P\big\{ a_1 P\{a_2\}\big\}(x) + P\big\{ P\{a_1\}a_2 \big\}(x).
$$
The functions $a_i=a_i(x)$, $i=1,2$ are defined over $\mathbb{R}$
and supposed to take values in a non-commutative algebra, say,
matrices of size $n \times n$. Then we find \allowdisplaybreaks{
\begin{eqnarray}
    \label{Magnus1}
    P\big\{\chi_0(a)\big\}(x) &=& P\{a\}(x) - \frac 12 P\big\{ [P\{a\},\,a] \big\}(x)
                            +  \frac 14 P\Big\{\big[P\big\{[P\{a\},\,a]\big\},\,a\big] \Big\}(x) \nonumber \\
                      & &\hspace{7cm} + \frac{1}{12} P\Big\{\big[P\{a\},\,[P\{a\},\,a]\big]\Big\}(x) + \cdots
\end{eqnarray}}
Let us write the terms in (\ref{Magnus1}) explicitly
\allowdisplaybreaks{
\begin{align}
  &  P\{a\}(x)=\int_0^{x}a(y)dy\\
  &  \frac 12 P\big\{ [P\{a\},\,a] \big\}(x) =  \frac 12 \int_0^{x}\int_0^{y_1} [ a(y_2),\,a(y_1)] dy_2\ dy_1\\
  &   \frac 14 P\Big\{\big[P\big\{[P\{a\},\,a]\big\},\,a\big] \Big\}(x)
        =  \frac 14 \int_0^{x} \int_0^{y_1}\int_0^{y_2} \big[ [ a(y_3),\,a(y_2)],\, a(y_1)\big] dy_3\ dy_2\ dy_1\\
  &  \frac{1}{12} P\Big\{\big[P\{a\},\,[P\{a\},\,a]\big]\Big\}(x) =
        \frac {1}{12} \int_0^{x} \int_0^{y_1}\int_0^{y_1} \big[ a(y_3),\,[ a(y_2),\, a(y_1)]\big] dy_3\ dy_2\ dy_1.
\end{align}}
Baxter's original motivation was to generalize the integral
equation
\begin{equation}
    \label{eq:integralIVP}
    f(x)=1+P\{fa\}(x)
\end{equation}
corresponding to the first order initial value problem
\begin{equation}
    \label{eq:IVP}
    \frac{d}{dx}f(x)=a(x)f(x),\quad f(0)=1
\end{equation}
with unique solution
\begin{equation}
    \label{exp-sol}
     f(x) = \exp\big(P\{a\}(x)\big)
\end{equation}
by replacing the Riemann integral by another Rota--Baxter map $P$
of non-zero weight $\theta$ (\ref{eq:RB}) on a commutative
algebra. The result is the classical Spitzer identity
(\ref{SpitzerId}), which in the more general non-zero weight
$\theta$ case takes the form \allowdisplaybreaks{
\begin{eqnarray}
  \exp\Big(- P\Big(\frac{\log(1 - \theta a)}{\theta}\Big) \Big)
         &=&\sum_{n=0}^\infty  \underbrace{P\big(P(P( \cdots P}_{n-times}(a)a)\dots a)a\big).
  \label{SpitzerId-theta}
\end{eqnarray}}
This follows form (\ref{id:NC-SpitzerId-theta}) with $b=-a$, since
$\chi_{\theta}=\id_A$ in the commutative case. One readily
verifies that the left-hand side of this identity reduces to the
exponential $\exp \big(P(a)\big)$, compare with (\ref{exp-sol}),
in the limit $\theta \to 0$.

To summarize, Proposition \ref{pp:BCHtheta} generalizes
Proposition \ref{pp:sp} to non-commutative weight $\theta \neq 0$
Rota--Baxter algebras. Corollary \ref{cor:NC-SpitzerId-theta}
describes an extension of Baxter's result on Spitzer's identity to
general associative Rota--Baxter algebras of weight $\theta
\neq0$, i.e., not necessarily commutative. The particular case of
vanishing weight $\theta \to 0$ is captured by the following

\begin{lem}\label{thm:0-ncSpitzer}
Let $(A,P)$ be a complete filtered Rota--Baxter algebra of weight
zero. For $a \in A_1$ the weight zero $BC\!H$-recursion $\chi_0
:A_1\to A_1$ is given by the recursion in Eq.~(\ref{BCH-recur5a})
\begin{equation*}
 \chi_0(a)=-\frac{\hbox{ad}P\big(\chi_0(a)\big)}{1-{\rm{e}}^{\hbox{\eightrm ad}P(\chi_0(a))}}(a).
\end{equation*}
\begin{enumerate}
\item\label{eq:exp1o} The equation $x=1-P(x\ a)$ has a
                      unique solution $x = \exp\big(-P(\chi_0(a))\big)$.
\item\label{eq:exp2o} The equation $y=1 + P(a\ y)$ has a unique
                      solution $y=\exp\big(P(\chi_0(a))\big).$
\end{enumerate}
\end{lem}
Atkinson's factorization for the weight zero case,
Eq.~(\ref{AtkinsonZero}), follows immediately from the preceding
lemma.

In view of example (\ref{Magnus1}) the last lemma leads to the
following corollary. Recall the work by Magnus~\cite{Mag} on
initial value problems of the above type but in a non-commutative
setting, e.g., for matrix-valued functions. He proposed an
exponential solution
$$
    F(x) = \exp\big(\Omega[a](x)\big)
$$
with $\Omega[a](0)=0$, for the  first order initial value problem
$\frac{d}{dx}F(x)=a(x)F(x)$, $F(0)=1$, respectively the
corresponding integral equation $F(x) = 1 + P\{aF\}(x)$, where $P$
is again, of course, the Riemann integral operator. He found an
expansion for $\Omega[a](x)=\sum_{n>0}\Omega^{(n)}[a](x)$ in terms
of multiple integrals of nested commutators, and provided a
recursive equation for the terms $\Omega^{(n)}[a](x)$:
\begin{equation}
 \frac{d}{dx}\Omega[a](x) = \frac{\hbox{ad}\,
                \Omega[a]}{{\rm{e}}^{\hbox{\eightrm ad}\,\Omega[a]}-1}(a)(x).
\end{equation}
Comparison with (\ref{BCH-recur5a}) (and also (\ref{BCH-recur5b})
and (\ref{pre-Magnus})) settles the link between Magnus recursion
and $BC\!H$-recursion in the context of a vanishing Rota--Baxter
weight, namely

\begin{cor} Let $A$ be a function algebra over $\mathbb{R}$ with
values in an operator algebra. $P$ denotes the indefinite Riemann
integral operator. Magnus' $\Omega$ expansion is given by the
formula
\begin{equation}
    \label{magnus-link}
    \Omega[a](x)=P\big(\chi_0(a)\big)(x).
\end{equation}
\end{cor}

Hence, the $\theta$-$BC\!H$-recursion (\ref{BCH-recursion3})
generalizes Magnus' expansion to general weight $\theta \neq 0$
Rota--Baxter operators $P$ by replacing the weight zero Riemann
integral in $F= 1 + P\{aF\}$.

The following commutative diagram (\ref{diag:general}) summarizes
the foregoing relations. Generalizing the simple initial value
problem in (\ref{eq:IVP}) twofold. First we go to the integral
equation in (\ref{eq:integralIVP}). Then we replace the Riemann
integral by a general Rota--Baxter map and assume a
non-commutative setting.

Hence, we start with a complete filtered non-commutative
associative Rota--Baxter algebra $(A,P)$ of non-zero weight
$\theta \in \mathbb{K}$. The top of (\ref{diag:general}) contains
the solution to the recursive equation
\begin{equation}
    \label{eq:recursion}
    y = 1 + P(y\ b)
\end{equation}
for $b \in A_1$ which is given in terms of Spitzer's identity
generalized to associative otherwise arbitrary Rota--Baxter
algebras (\ref{id:NC-SpitzerId-theta}),
\begin{equation}
    \label{eq:solution}
    y=\exp\Big(- P\Big(\chi_{\theta}\Big(\frac{\log(1 - \theta b)}{\theta}\Big)\Big)\Big).
\end{equation}
The $\theta$-$BC\!H$-recursion $\chi_{\theta}$ is given in
(\ref{BCH-recursion4}). The left wing of (\ref{diag:general})
describes the case when first, the weight $\theta$ goes to zero,
hence reducing $\chi_\theta \to \chi_0$. This is the algebraic
structure underlying Magnus' $\Omega$-expansion. Then the algebra
$A$ becomes commutative which implies $\chi_0 = \id_A$. The right
wing of diagram (\ref{diag:general}) just describes the opposite
reduction, i.e., we fist make the algebra commutative, which gives
the classical Spitzer identity for non-zero weight commutative
Rota--Baxter algebras (\ref{SpitzerId-theta}). Then we take the
limit $\theta \to 0$.
\begin{equation}
    \label{diag:general}
    \xymatrix{
                        & {\hbox{{\eightrm{$\exp\!\Big(\!\!-\!\! P\Big(\chi_{\theta}
                                                    \Big(\frac{\log(1 - \theta
                                                    b)}{\theta}\Big)\!\!\Big)\!\Big)$}}}
                                                    \atop \theta \neq 0,\ non-com. }
                                                     \ar[dd]^{{com.\atop \theta \to 0}}
                                                     \ar[rd]_{\theta \neq 0 \atop com.}
                                                     \ar[ld]^{\theta \to 0 \atop {non-com.}}
                                                     &\\
   {\hbox{\eightrm{$\exp\!\big(P\big(\chi_0(b)\big)\!\big)$}}
    \atop {\rm{Magnus}}}
   \ar[rd]^{com.}   &
                                                    & {\hbox{{\eightrm{$\exp\!\Big(\!\!-\!\!P\Big(
                                                     \!\frac{\log(1 - \theta b)}{\theta}
                                                     \!\Big)\!\Big)$}}}
                                                      \atop {\rm{cl.\ Spitzer}}}
                                                      \ar[ld]_{\theta \to 0}\\
                   & {\hbox{\eightrm{$\exp\!\big(P(b)\big)$}}
                      \atop \theta = 0,\ com.}&
                }
\end{equation}
Both paths eventually arrive at the simple fact that equation
(\ref{eq:recursion}) is solved by a simple exponential in a
commutative weight zero Rota--Baxter setting. This is the general
algebraic structure underlying the the initial value problem in
(\ref{eq:IVP}) respectively its corresponding integral equation
(\ref{eq:integralIVP}).


\section{Renormalization in perturbative QFT}
\label{sect:CKqft}

This section recalls some of the results
from~\cite{E-G-K2,E-G-K3}. We derive Connes' and Kreimer's
Birkhoff decomposition of Hopf algebra characters with values in a
commutative unital Rota--Baxter algebra. For more details we refer
the reader to \cite{EG,EK,FG,Ma}.

In most of the interesting and relevant 4-dimensional quantum
field theories (QFT), to perform even simple perturbative
calculations, one can not avoid facing ill-defined integrals. The
removal of these (ultraviolet) divergencies, or short-distance
singularities, in a physically and mathematically sound way is the
process of renormalization \cite{Collins84}.

In the theory of Kreimer \cite{KreimerHopf}, and Connes and
Kreimer Feynman graphs as the main building blocks of perturbative
QFT are organized into a Hopf algebra. In particular, Connes and
Kreimer discovered a Birkhoff type decomposition for Hopf algebra
characters with values in the field of Laurent series, which
captures the process of renormalization. We will briefly outline
an algebraic framework for this decomposition based on the above
results of Spitzer and Atkinson.

We work in the setting of Connes and Kreimer~\cite{CK2}. Recall
that in the language of Kreimer for a given perturbative
renormalizable QFT, denoted by $\mathcal{F}$, we have a graded,
connected, commutative, non-cocommutative Hopf algebra
$\mathcal{H}_\mathcal{F} := (\mathcal{H} :=\bigoplus_{n\geq 0}
\mathcal{H}_n, \Delta, m_\mathcal{H}, \varepsilon_\mathcal{H},S)$
of one-particle irreducible (1PI) Feynman graphs with coproduct
$\Delta$ defined by
$$
    \Delta(\Gamma) = \Gamma \otimes 1 + 1 \otimes \Gamma
                                +\sum_{\gamma\subset \Gamma} \gamma \otimes \Gamma / \gamma.
$$
Here the sum is over all 1PI ultraviolet divergent subgraphs
$\gamma$ in $\Gamma$ and $\Gamma/\gamma$ denotes the corresponding
cograph. The decomposition of $\Gamma$ in $\Delta(\Gamma)$
essentially describes the combinatorics of renormalization.

The space $\Hom(\mathcal{H}_\mathcal{F},\CC)$ of linear maps
$\mathcal{H}_\mathcal{F} \to \CC$ equipped with the convolution
product $f \star g:=m_\mathcal{H}\circ(f \otimes g)\circ\Delta$ is
an associative algebra with the counit $\varepsilon_\mathcal{H}$
as unit. $\Hom(\mathcal{H}_\mathcal{F},\CC)$ contains the group
$G: = \mchar(\mathcal{H}_\mathcal{F},\CC)$ of Hopf algebra
characters, i.e., algebra homomorphisms, and its corresponding Lie
algebra $\frakg := \frakg_\mathcal{F} =
\partial\mchar(\mathcal{H}_\mathcal{F},\CC)$ of derivations
(infinitesimal characters). Feynman rules in $\mathcal{F}$ provide
such an algebra homomorphism from $\mathcal{H}_\mathcal{F}$ to
$\CC$.

In general, ultraviolet divergencies demand a regularization
prescription, where by introducing extra parameters, the
characters become algebra homomorphisms, say for instance, into $L
= \CC[\ep^{-1},\ep]]$, the field of Laurent series (dimensional
regularization scheme). We denote by $G_L :=
\mchar(\mathcal{H}_\mathcal{F},L) \subset
\Hom(\mathcal{H}_\mathcal{F},L)$ the group of $L$-valued, or
regularized, algebra homomorphisms. Hence, now the set of Feynman
rules together with dimensional regularization amounts to a linear
map from the set of 1PI Feynman graphs to $L$ and hence an algebra
homomorphism denoted by $\phi\in G_L$ from
$\mathcal{H}_\mathcal{F}$ to $L$.

We will now make the connection to
subsection~\ref{subsect:Rota-Baxter}. The field of Laurent series
actually forms a commutative Rota--Baxter algebra $(L,R)$ with the
projector $R$ on $L$
$$
    R:L \to L,\; \sum_{i=-n}^\infty a_i \ep^i\mapsto \sum_{i=-n}^{-1}a_i \ep^i
$$
to the strict pole part of a Laurent series as the idempotent
weight-one Rota--Baxter map (minimal subtraction scheme).

It was shown in \cite{CK2,KreimerHopf} that this setup allows for
a concise Hopf algebraic description of the process of
perturbative renormalization of the QFT $\mathcal{F}$. To wit,
Connes and Kreimer observed that Bogoliubov's recursive formula
for the counter term in renormalization has a Hopf algebraic
expression given by inductively defining the map $\phi_- \in
\mathcal{H}_\mathcal{F}$
\begin{equation}
    \label{eq:CK-ct}
    \phi_-(\Gamma)=-R\big(\phi(\Gamma)
                + \sum_{\gamma\subset \Gamma} \phi_-(\gamma)\phi(\Gamma / \gamma)\big)
\end{equation}
with $\phi_-(\Gamma)=-R(\Gamma)$ if $\Gamma$ is a primitive
element in $\mathcal{H}_\mathcal{F}$, i.e., contains no
subdivergence. The map
$$
    \bar{{\rm{R}}}[\phi](\Gamma):=\phi(\Gamma) +
                    \sum_{\gamma\subset \Gamma} \phi_-(\gamma)\phi(\Gamma / \gamma)
$$
for $\Gamma \in \ker(\vep_\mathcal{H})$ is Bogoliubov's
preparation map. This lead to the Birkhoff decomposition of
Feynman rules found by Connes and Kreimer
\cite{CK1,CK2,CK3,KreimerChen}, described in the following
theorem.

\begin{thm} \label{thm:ck}
The renormalization of $\phi \xrightarrow{ren.} \phi_+$ follows
from the convolution product of the counter term $\phi_-$
(\ref{eq:CK-ct}) with $\phi$, $\phi_+:=\phi_-\star \phi$, implying
the inductive formula for $\phi_+$
$$
    \phi_{+}(\Gamma) = \phi(\Gamma) + \phi_-(\Gamma)+
           \sum_{\gamma\subset \Gamma} \phi_-(\gamma)\phi(\Gamma / \gamma).
$$
Further, the maps $\phi_-$ and $\phi_+$ are the unique characters
such that $\phi = \phi_-^{-1} \star \phi_+$ gives the algebraic
Birkhoff decomposition of the regularized Feynman rules character
$\phi \in G_{L}$.
\end{thm}

The following theorem describes the Birkhoff decomposition of
Connes and Kreimer in Theorem \ref{thm:ck} using the algebraic
setting developed in the earlier sections.

\begin{thm}\label{thm:BCH-RB-QFT}{\rm{\cite{E-G-K2,E-G-K3}}}
In Proposition \ref{pp:sp}, take $A$ to be
$(\Hom(\mathcal{H}_\mathcal{F},L),\mathcal{R})$, which is a
complete filtered Rota--Baxter algebra with Rota--Baxter operator
$\mathcal{R}(\phi):= R \circ \phi$ and filtration from
$\mathcal{H}$. We denote its unit by $e:= u_L \circ
\varepsilon_{\mathcal{H}}$. For a $L$-valued character $\phi \in
\mchar(\mathcal{H}_\mathcal{F},L)$ take $b:=\phi - e$. Then one
can show that $b \in A_1$ and

\begin{enumerate}
\item
    the equations in (\ref{eq:recurs}) are the recursive formulae for
    $x =: \phi_-$ and $x' =:\phi_+^{-1}$ in the work of Connes--Kreimer;

\item
    the exponential factors in Equation~(\ref{eq:bch}) give the unique
    explicit formulae for $x^{-1}=\phi_-^{-1}$ and $x'^{-1}=\phi_+$.

\item
    equation (\ref{eq:bch}) gives the unique Birkhoff
    decomposition of  $\phi = \phi_-^{-1} \star \phi_+$
    found in Connes--Kreimer's work;

\item
    Bogoliubov's $\bar{{\rm{R}}}$-map$, \bar{{\rm{R}}}[\phi]: \mathcal{H} \to
    L$,  is given by $\bar{{\rm{R}}}[\phi] = \exp^{\star_\mathcal{R}}\big(-\chi(\log(\phi))\big)$,
    for $\phi \in G_L$, such that $\tilde{\mathcal{R}}(\bar{{\rm{R}}}[\phi])
    =2e-\phi_{+}$ and $\mathcal{R}(\bar{{\rm{R}}}[\phi])
    =\phi_{-} - e$.
    Here $\phi_1{\star_\mathcal{R}}\phi_2:=\mathcal{R}(\phi_1)\star\phi_2 + \phi_1\star\mathcal{R}(\phi_2) - \phi_1 \star
    \phi_2$, $\phi_i\in G_L$, $i=1,2$, see (\ref{doubleRB}).
\end{enumerate}
\end{thm}

It is evident that one can replace the particular choice of the
field of Laurent series $L$ by any other commutative Rota--Baxter
algebra with idempotent Rota--Baxter map. Proposition \ref{pp:sp}
provides us with a recursion for the renormalization of $\phi
\xrightarrow{ren.} \phi_+$ which does not contain the counter term
$\phi_{-}$.

\begin{cor} \cite{E-G-K2,EG,EGGV}
With the assumption of Theorem~\ref{thm:BCH-RB-QFT}, the second
equation in (\ref{eq:inverseRecurs1}) gives a recursion for
$\phi_+$
$$
    \phi_{+} = e -
    \mathcal{\tilde{R}}\big(\phi_{+}\star(\phi^{-1} - e)\big).
$$
\end{cor}
\noindent Recall that the inverse of $\phi \in G_L$ is given by
the composition with the antipode, $\phi^{-1}=\phi \circ S$.\\

We should mention that in recent work~\cite{EGGV,EG} the first two
authors showed, together with J.~M.~Gracia-Bond\'{i}a and
J.~C.~V\'{a}rilly, how the combinatorics of perturbative
renormalization can be represented by matrix factorization of
unipotent upper (lower) triangular matrices with entries in a
commutative Rota--Baxter algebra. As we have seen above such
triangular matrices provide a simple example of a complete
filtered Rota--Baxter algebra.


\section{Formal exponentials}
\label{sect:BHL}

In this section we consider now the factorization of formal
exponentials described by Barron, Huang and Lepowsky in \cite{B-H-L}
in the context of the $BC\!H$-recursion map $\chi$. Let us first
recall their notations and results.

Let $\frakg$ be a Lie algebra with a decomposition $\frakg^-\oplus
\frakg^+$ of the underlying vector space. Equivalently, there is
an idempotent linear map $P:\frakg \to \frakg$. Then we have the
corresponding decomposition of the complete Lie algebra
$$
    \frakg[[s,t]] = \frakg^-[[s,t]] \oplus \frakg^+[[s,t]].
$$
Consider the (restriction of the) $\BCH$ map
$$
    C: \frakg[[s,t]]_1 \times \frakg[[s,t]]_1 \to \frakg[[s,t]]_1.
$$
Here $\frakg[[s,t]]_1 = s\frakg[[s,t]] + t\frakg[[s,t]]$.

\begin{thm} \label{thm:bhl} \cite{B-H-L}
\begin{enumerate}
    \item \label{bhl1} {\rm{(Factorization Theorem of Barron--Huang--Lepowsky)}} The map
        $$
            C:s\frakg^-[[s,t]]\times t\frakg^+[[s,t]] \to s\frakg^-[[s,t]]\oplus t\frakg^+[[s,t]]
        $$
        is bijective. Here we have direct product on the source space and (direct) sum in
        $\frakg[[s,t]]$ on the target space.

    \item \label{bhl2} {\rm{(Formal Algebraic Uniformization of Barron--Huang--Lepowsky)}} There
            exists a unique bijection
        $$
            \Psi=(\Psi_-,\Psi_+): t\frakg^+[[s,t]] \times s\frakg^-[[s,t]] \to s\frakg^-[[s,t]] \times t\frakg^+[[s,t]]
        $$
        such that for $g^\pm\in \frakg^\pm[[s,t]]$,
        $$
            \exp(tg^+) \exp(sg^-) = \exp\Psi_-(tg^+,sg^-) \exp\Psi_+(tg^+,sg^-).
        $$
\end{enumerate}
\end{thm}

They further posed the following problem.

\begin{problem} \label{pr:bhl} \cite[Problem 3.2]{B-H-L}
Find a closed form for the inverse map of $C$ in
Theorem~\ref{thm:bhl}.(\ref{bhl1}).
\end{problem}

The theorem of Barron, Huang and Lepowsky generalizes a well-known
result in the case when $\frakg$ is a finite-dimensional Lie
algebra over $\RR$ or $\CC$. In this special case, the proof was
obtained by a geometric argument on the corresponding Lie group.
Actually, their proof is algebraic, making use of a more precise
expression of $C(x,y)$ and the gradings given by $s$ and $t$. We
will show here how to derive this result from
Theorem~\ref{thm:idemp}.

\begin{thm} \label{thm:bhl2}
\begin{enumerate}
    \item \label{factor1} The bijection $C$ in Theorem~\ref{thm:bhl}.(\ref{bhl1}) is the restriction $C_-$ of the
          bijection $C$ in part (\ref{it:factor}) of Theorem~\ref{thm:idemp}.

    \item \label{uniform1} The uniformization $\Psi$ in Theorem~\ref{thm:bhl}.(\ref{bhl2}) is the restriction
          of the uniformization $\Psi$ in Theorem~\ref{thm:idemp}.(\ref{it:uniform}).

    \item \label{closed1} The restriction of the formulae for $C_-$ and
          $\Psi$ in Theorem~\ref{thm:idemp}.(\ref{it:closed}) gives the formulae for
          the inverse map of $C$ and $\Psi$ in Theorem~\ref{thm:bhl}.(\ref{bhl1}).
\end{enumerate}
\end{thm}

\begin{remark}{\rm{
The formulae in item (\ref{closed1}) involve the $BC\!H$-recursion
$\chi$ which is defined in terms of the recursive equation
(\ref{BCHrecursion1}), respectively (\ref{BCHrecursion2}). Hence,
our approach allows for a compact formulation of
Problem~\ref{pr:bhl} in a generalized setting, to wit, find a
closed form for the recursively defined map $\chi$. Moreover,
Lemma~\ref{lem:algIdemRB} and especially
Equation~(\ref{chiSimple2}) of Section~\ref{sect:ABS} give
solutions to the $\BCH$-recursion $\chi$, that is, closed forms
for the inverse map of $C$, some particular situation.}}
\end{remark}

\begin{proof}
(\ref{factor1}) We first note that $\bar{\frakg}:=\frakg[[s,t]]$
is a complete Lie algebra with filtration defined by the grading
given by the total degree in the parameters $s$ and $t$. In
particular $\bar{\frakg}_1 = s\frakg[[s,t]] + t\frakg[[s,t]]$.
Thus by Theorem~\ref{thm:idemp}, the map
$$
    C_-: \frakg^-[[s,t]]_1 \times \frakg^+[[s,t]]_1\to \frakg[[s,t]]_1
        =\frakg^-[[s,t]]_1 \oplus \frakg^+[[s,t]]_1$$
is bijective with inverse
$$
    D_P=(P,\ \tilde{P}) \circ \chi:
    \frakg^-[[s,t]]_1 \oplus \frakg^+[[s,t]]_1 \to \frakg^-[[s,t]]_1 \times \frakg^+[[s,t]]_1.
$$
Then to prove items~(\ref{uniform1}) and~(\ref{closed1}) in
Theorem~\ref{thm:bhl2}, and hence Theorem~\ref{thm:bhl}, we only
need to show

\begin{lem}
Let $U = s\frakg^-[[s,t]] \times t\frakg^+[[s,t]]$, $V =
s\frakg^-[[s,t]] \oplus t\frakg^+[[s,t]].$ Then $C_-$ restricts to
a bijective map from $U$ onto $V$.
\end{lem}

\begin{proof}
The inclusion $C(U)\subset V$ is straightforward: if $a_-\in
s\frakg^-[[s,t]]\times t\frakg^+[[s,t]]$ then clearly $a_- + a_+
\in V$, and $\BCH(a_-,a_+)\in st\frakg[[s,t]] =
st\frakg^-[[s,t]]\oplus st\frakg^+[[s,t]]\subseteq
s\frakg^-[[s,t]]\oplus t\frakg^+[[s,t]]$.

Let us now prove the inclusion $\chi(V)\subseteq V$, i.e.,
\begin{equation}
    \chi\big(s\frakg^-[[s,t]]\oplus t\frakg^+[[s,t]]\big) \subseteq s\frakg^-[[s,t]]\oplus t\frakg^+[[s,t]],
    \label{eq:inc}
\end{equation}
by using the definition of $BC\!H$-recursion: for any $v \in V$,
$\chi(v)$ is the limit of $\chi_{(n)}(v)$ (for the topology
defined by the filtration) recursively defined by
$\chi_{(0)}(v)=v$ and
$$
    \chi_{(n)}(v)=v - \BCH\left(P\big(\chi_{(n-1)}(v)\big),\,\tilde P\big(\chi_{(n-1)}(v)\big)\right).
$$
It is clear, from the same argument as above, that
$\chi_{(n-1)}(v)\in V$ implies $\chi_{(n)}(v)\in V$, so
$\chi_{(n)}(v)\in V$ for any $n$ by induction. We then deduce
$\chi(v) \in V$ by taking the limit, as $V$ is closed. We deduce
immediately from this inclusion that $D_P(V)\subseteq U$, as
$D_P(a)=\big(P\circ\chi(a),\,\tilde P\circ\chi(a)\big).$
\end{proof}

\noindent (\ref{uniform1}) Recall that the map $\Psi$ in
Theorem~\ref{thm:bhl}.(\ref{bhl2}) is given by the following
diagram
$$
    \xymatrix{ t\frakg^+[[s,t]]\times s\frakg^-[[s,t]] \ar^{C}[rr]
                \ar^{\Psi}[d] && t\frakg^+[[s,t]]\oplus s\frakg^-[[s,t]] \ar_{\sigma}[d]\\
                    s\frakg^-[[s,t]]\times t\frakg^+[[s,t]] && \ar^{D}[ll]
                    s\frakg^-[[s,t]]\oplus t\frakg^+[[s,t]]
            }
$$
Here again $\sigma$ is just the identity map
$$
    \sigma(t\,h^+ + s\,h^-)=s\,h^-+ t\,h^+.
$$
Then the proof of part (\ref{uniform1}) follows from part
(\ref{factor1}).

\noindent Part (\ref{closed1}) is readily verified.
\end{proof}


\section{Combinatorial Hopf algebras}
\label{sect:ABS}

In Section \ref{sect:CKqft} we applied the factorization property
of the $BC\!H$-recursion $\chi$ together with the Rota--Baxter
relation to Hopf algebras, in the context of the Hopf algebraic
description of renormalization by Connes and Kreimer. This section
consists of another application of $\chi$ to connected graded Hopf
algebras. We analyze explicitly the even-odd decomposition of
Aguiar, Bergeron, and Sottile~\cite{A-B-S}\footnote{We thank
W.~Schmitt for bringing the paper of Aguiar, Bergeron and Sottile
\cite{A-B-S} to our attention.}.

For a connected graded Hopf algebra $(\mathcal{H}=\oplus_{n\geq 0}
\mathcal{H}_n,\Delta,m,\varepsilon,S)$, we define the grading
operator $Y(h):= |h|h :=nh$, for a homogeneous element $h \in
\mathcal{H}_n$, and extend linearly.

The grading on $\mathcal{H}$ defines a canonical involutive
automorphism on $\mathcal{H}$, denoted by
$\phantom{i}\overline{\phantom{a}}: \mathcal{H} \to \mathcal{H}$,
$\overline{h} := (-1)^{|h|}h = (-1)^{n} h$, for $h \in
\mathcal{H}_n$. It induces by duality an involution on
$\Hom(\mathcal{H},\mathbb{K})$, $\overline{\phi}(h) =
\phi(\bar{h})$ for $\phi \in \Hom(\mathcal{H},\mathbb{K})$, $h \in
\mathcal{H}$.

$\mathcal{H}$ naturally decomposes into $\mathcal{H}_{-} :=
\bigoplus_{n>0} \mathcal{H}_{2n-1}$ and $\mathcal{H}_{+}:=
\bigoplus_{n \geq 0} \mathcal{H}_{2n}$ on the level of vector
spaces
$$
  \mathcal{H}=\mathcal{H}_{-} \oplus \mathcal{H}_{+},
$$
with projectors $\pi_{\pm}: \mathcal{H} \to \mathcal{H}_{\pm}$.
Such that for $h \in \mathcal{H}$, $\overline{\pi_{+}(h)} =
\overline{h_+}=h_+$ and $\overline{\pi_{-}(h)} = \overline{h_-}=
-h_-$, $h = h_- + h_+$. As a remark we mention that
$\mathcal{H}_{+}$ is a subalgebra of $\mathcal{H}$, whereas
$\mathcal{H}_{-}$ is just a subspace, hence neither $\pi_{-}$ nor
$\pi_{+}:=\id_\mathcal{H} - \pi_{-}$ are Rota--Baxter maps.
Instead, we have $\mathcal{H}_{\pm}\mathcal{H}_{\pm} \subset
\mathcal{H}_{+}$ and $\mathcal{H}_{\pm}\mathcal{H}_{\mp} \subset
\mathcal{H}_{-}$.

The set of characters $G:=\mchar(\mathcal{H},\mathbb{K})$, i.e.,
multiplicative maps $\phi \in \Hom(\mathcal{H},\mathbb{K})$, forms
a group under convolution, defined by
$$
    f \star g:= m_{\mathbb{K}}\circ(f \otimes g)\circ \Delta,
$$
for $f,g \in \Hom(\mathcal{H},\mathbb{K})$. A character $\phi \in
G$ is called even if it is a fixed point of the involution,
$\overline{\phi} =\phi$, and is called odd if it is an anti-fixed
point, $\overline{\phi}=\phi^{-1}=\phi \circ S$. The set of odd
and even characters is denoted by $G_-$, $G_+$, respectively. Even
characters form a subgroup in $G$. Whereas the set of odd
characters forms a symmetric space. The following theorem is
proved in \cite{A-B-S}.

\begin{thm} {\rm{\cite{A-B-S}}}
Any $\phi \in \mchar(\mathcal{H},\mathbb{K})$ has a unique
decomposition $\phi = \phi_- \star \phi_+$ with $\phi_- \in G_-$
being an odd character, and $\phi_+ \in G_+$ being an even
character. \label{thm:abs}
\end{thm}

Both projectors $\pi_-: \mathcal{H} \to \mathcal{H}_-$ and $\pi_+:
\mathcal{H} \to \mathcal{H}_+$ lift to
$\Hom(\mathcal{H},\mathbb{K})$. Implying for the complete filtered
Lie algebra $\frakg:=\partial \mchar(\mathcal{H},\mathbb{K})$,
with filtration from $\mathcal{H}$, the direct decomposition
$\frakg = \frakg_- \oplus \frakg_+$ into the Lie subalgebra
$\frakg_+$ and the Lie triple system $\frakg_-$. Such that for any
$Z \in \frakg$, we have $Z = Z_- + Z_+$, $Z_{\pm}\in \frakg_{\pm}$
unique. Then by Theorem~\ref{thm:bch}, there is a
$BC\!H$-recursion, $\chi: \frakg_1 \to \frakg_1$ such that, for
any $\phi = \exp(Z) \in \mchar(\mathcal{H},\mathbb{K})$, $Z \in
\frakg_1$, we have
\begin{equation} \label{eq:bchdecomp}
    \phi = \exp(Z) = \exp\big(Z_- + Z_+\big)
         = \exp \big(\chi(Z)_-\big) \star \exp\big(\chi(Z)_+\big).
\end{equation}
Here the exponential is defined with respect to the convolution
product, $\exp(Z):=\sum_{n \geq 0} \frac{Z^{\star\ n}}{n!}$, but
we will skip the $\star$ in the following to ease the notation.

\begin{thm} \label{thm:ABS2}
The even-odd factorization of a character in Theorem~\ref{thm:abs}
coincides with the factorization in item (\ref{it1:bch-thm}) of
Theorem~\ref{thm:bch}.
\end{thm}

The proof follows from the following properties of the involution
$\overline{\phantom{a}}: \mathcal{H} \to \mathcal{H}$. Recall the
definition of an algebra involution on an algebra $A$, which is an
algebra homomorphism $\jmath: A \to A$ such that $\jmath^2=\id_A$.
Dually, define now a {\bf{coalgebra involution}} to be a linear
map $\jmath$ on a coalgebra $C$ such that $\jmath^2=\id_C$ and
$(\jmath \otimes \jmath) \circ \Delta = \Delta \circ \jmath$.

\begin{lem}\label{lem:inv}
Let $\mathcal{H}$ be a connected filtered Hopf
$\mathbb{K}$-algebra. Let $\jmath: \mathcal{H} \to \mathcal{H}$ be
a coalgebra involution preserving the filtration. Then by
pre-composition, $\jmath$ defines an algebra involution, still
denoted by $\jmath$, on the filtered algebra
$\mathcal{A}:=\Hom(\mathcal{H},\mathbb{K})$ that preserves the
filtration.
\end{lem}

\begin{proof}
For $h \in \mathcal{H}$, we have
\begin{eqnarray*}
    \jmath(f\star g)(h)=(f\star g)(\jmath(h))&=&(m_{\mathbb{K}}\circ (f\otimes g)\circ \Delta\circ \jmath)(h)\\
                                             &=&(m_{\mathbb{K}}\circ (f\otimes g)\circ (\jmath\otimes \jmath)\circ \Delta)(h)\\
                                             &=&(\jmath(f)\star \jmath(g))(h).
\end{eqnarray*}
So $\jmath: \mathcal{A} \to \mathcal{A}$ is an algebra
homomorphism. Clearly, $\jmath$ preserves the filtration and
$\jmath^2=\id$.
\end{proof}

\begin{lem} \label{lem:invexp}
Let $\jmath$ be an algebra involution on the complete filtered
algebra $\mathcal{A}$ that preserves the filtration.
\begin{enumerate}
    \item \label{it:exp1} If $\jmath(a) = \pm a$ for $a \in \mathcal{A}_1$,
    then $\jmath(\exp (a)) = \exp(\pm a)$.

    \item \label{it:exp2} Let
    $\mathcal{A}_{1,\pm}:= \{a\in \mathcal{A}_1\, \big|\, \jmath(a)=\pm a\}$ and
    $G_\pm := \{ \eta \in 1 + \mathcal{A}_1\, \big|\, \jmath(\eta) = \eta^{\pm 1}\}.$
    Then $\exp\big(\mathcal{A}_{1,-}\big) = G_-$ and $\exp\big(\mathcal{A}_{1,+}\big) = G_+$.
\end{enumerate}
\end{lem}

\begin{proof}
Since $\jmath$ preserves the filtration, $\jmath$ is a continuous
map with respect to the topology defined by the filtration. So for
any $a \in \mathcal{A}_1$, we have
$$
    \jmath \big(\exp (a)\big)=\jmath \Big(\lim_{k \to \infty} \sum_{n=0}^k \frac{a^n}{n!}\Big)
                    = \lim_{k\to \infty} \jmath \Big(\sum_{n=0}^k \frac{a^n}{n!}\Big)
                    = \lim_{k\to \infty} \sum_{n=0}^k \frac{\jmath(a)^n}{n!}
                    = \exp(\jmath(a)).
$$
Now item (\ref{it:exp1}) of the lemma follows. Item
(\ref{it:exp2}) then follows from the bijectivity of $\exp$.
\end{proof}

\begin{proof}
(of Theorem~\ref{thm:ABS2}) Now let $\phi \in
{\rm{Char}}(\mathcal{H},\mathbb{K})$, and $\jmath = \bar{\ }$.
Then by Lemma~\ref{lem:inv}, we see that the induced $\jmath =
\bar{\ }$ on $\Hom(\mathcal{H},\mathbb{K})$ is an algebra
involution that preserves the filtration. Then by
Lemma~\ref{lem:invexp}.(\ref{it:exp1}),
$\exp\big(\chi(Z)_{-}\big)$ (resp. $\exp\big(\chi(Z)_{+}\big)$) is
odd (resp. even). So Eq.~(\ref{eq:bchdecomp}) gives a
decomposition of $\phi$ as an element of $G_-$ and $G_+$. By
Corollary~\ref{decomposition} and
Lemma~\ref{lem:invexp}.(\ref{it:exp2}), we must have
$\phi_{-}=\exp\big(\chi(Z)_{-}\big)$ and
$\phi_{+}=\exp\big(\chi(Z)_{+}\big)$, as needed.
\end{proof}

We should remind the reader that the results of Proposition
\ref{pp:sp} do not apply here. We cannot calculate the
exponentials $\phi_{\pm}$ using Spitzer's recursions in
(\ref{eq:recurs}), since neither the projector $\pi_{+}$ nor
$\pi_{-}$ are of Rota--Baxter type. Nevertheless, the particular
setting allows for a significant simplification of the
$BC\!H$-recursion. In fact, using that $\phi_{\pm} \in G_{\pm}$
hence $\overline{\phi_{-}}=\phi_{-}^{-1}$ and
$\overline{\phi_{+}}=\phi_{+}$ and some algebra \cite{Za1}, we
find the following simple formula for
$\pi_{-}(\chi(Z))=\chi(Z)_{-}$
\begin{equation}
    \label{chiSimple1}
    \pi_{-}(\chi(Z)) = \pi_{-}(Z) + \frac{1}{2} \BCH\big(\pi_{-}(Z) + \pi_{+}(Z), -\pi_{-}(Z)+\pi_{+}(Z)\big).
\end{equation}
This follows from Lemma~\ref{lem:invexp} implying
$\overline{\phi}=\jmath(\exp(Z))=\exp\big(-\pi_{-}(Z) +
\pi_{+}(Z)\big)$ but also
$$
   \jmath(\exp(Z))=\exp\big(-\chi(Z)_{-}\big) \star \exp\big(\chi(Z)_{+}\big).
$$
Therefore, we have $\phi \star
\overline{\phi}^{-1}=\phi_{-}\star\phi_{-}$ which gives
Eq.~(\ref{chiSimple1}). From the factorization in
Theorem~\ref{thm:ABS2} we derive a closed form for the
$BC\!H$-recursion
\begin{equation}
    \label{chiSimple2}
    \chi(Z) = Z + \BCH\Big( -\pi_{-}(Z) - \frac{1}{2} \BCH\big(Z, Z-2\pi_{-}(Z)\big) , Z \Big).
\end{equation}
We may remark that this gives an answer to Problem~\ref{pr:bhl} in
the particular setting just outlined, see also the remark after
Theorem~\ref{thm:bhl2}.


\section{Polar decomposition}
\label{sect:Zanna}

The factorization of Aguiar et al. in the context of connected
graded Hopf algebras is related to a general result elaborated in
more detail in \cite{Za1,Za2} and \cite{Za3}. There it is shown
that any connected Lie group $G$, together with an involutive
automorphism $\sigma$ on $G$ allows locally for a decomposition
similar to the above one.

We will briefly outline the setting of \cite{Za1,Za2,Za3} and show
that our $BC\!H$-recursion provides an efficient mean for
calculations. We should stress that the $BC\!H$-recursion approach
gives only formal series. Let $G$ be a connected Lie group, and
$\frakg$ its corresponding Lie algebra. We assume the existence of
an involutive automorphism $\sigma$ on $G$. Let $G_- := \{ \psi
\in G \ |\: \sigma(\psi) = \psi^{-1} \}$ denote the symmetric
space of anti-fixed points of $\sigma$, and by $G_+ := \{ \psi \in
G\ |\: \sigma(\psi) = \psi\}$ we denote the subgroup of fixed
points of $\sigma$. Also we denote by $\pi_{\pm}$ the lifted
projections on $\frakg$ corresponding to $\sigma$. Hence for the
Lie algebra $\frakg$ we have the direct decomposition in terms of
the images of these projectors, $\frakg=\frakg_{-} \oplus
\frakg_{+}$, where $\pi_{+}(\frakg)=:\frakg_{+}$ is a Lie
subalgebra and $\pi_{-}(\frakg)=:\frakg_{-}$ a Lie triple system.

In this setting Munthe-Kaas et al. derive a differentiable
factorization of $\psi = \exp(tZ) \in G$, $Z \in \frakg$ for
sufficiently small parameter $t$. Using the additive decomposition
of $\frakg$ in terms of the projectors $\pi_{\pm}$ corresponding
to $\sigma$, the factors, $\psi_{\pm}(t) := \exp\big(
X_{\pm}(Z;t)\big)$ in $\psi(t) =\psi_{-}(t)\psi_{+}(t)$ are
calculated solving differential equations in $t$. This way
explicit complicated recursions are derived for the terms in
$X_{\pm}(Z;t) = \sum_{i > 0} X^{(i)}_{\pm}(Z)\ t^i$, using the
relations between the spaces $\frakg_{\pm}$, i.e.,
\allowdisplaybreaks{
\begin{eqnarray}
    [ \frakg_{\pm},\frakg_{\mp} ] \subset \frakg_{-}, \qquad
    [ \frakg_{\pm},\frakg_{\pm} ] \subset \frakg_{+}. \label{rel2}
\end{eqnarray}}
The results coincide with those following form the simpler
$BC\!H$-recursion map $\chi$ (\ref{BCHrecursion1}), which we state
here again \allowdisplaybreaks{
\begin{equation*}
    \chi(Z) = Z - \BCH\big(\pi_{-}(\chi(Z)),\pi_{+}(\chi(Z))\big),
\end{equation*}}
or Eq.~(\ref{BCHrecursion2}) respectively its simple closed form
in Eq.~(\ref{chiSimple2}). The $\pi_{\pm}$ projections of the
first three terms of $\chi(Zt) = t \sum_{k \geq 0}
\chi^{(k)}(Z)t^k$ are $\pi_{\pm}(\chi^{(0)}(Z))=Z_{\pm}$ and for
the next two non-trivial parts (\ref{chi1},\ref{chi2}) we find in
order $t^2$

$$
 \pi_{+}\big( \chi^{(1)} (Z)\big) = X^{(2)}_{+} = 0 \qquad
 \pi_{-}\big( \chi^{(1)} (Z)\big) = X^{(2)}_{-} = -\frac{1}{2}[Z_-,Z_+]
$$

\noindent and the even respectively odd projections in order $t^3$

\allowdisplaybreaks{
\begin{eqnarray}
 \pi_{+}\big(\chi^{(2)}(Z)\big) &=& X^{(3)}_{+} =\frac{1}{12}\big[Z_-,[Z_-,Z_+]\big],\\
 \pi_{-}\big(\chi^{(2)}(Z)\big) &=&  X^{(3)}_{-} =  -\frac{1}{4}[Z_+,[Z_-,Z_+]]+\frac{1}{12}[Z_+,[Z_-,Z_+]] \nonumber\\
                                &=& -\frac{1}{6}[Z_+,[Z_-,Z_+]].
\end{eqnarray}}

\noindent The reader is invited to compare them with the results
in Munthe-Kaas et al.~\cite{Za1,Za2} and especially Zanna's work
\cite{Za3}\footnote{We would like to point to the recursive
equation (1.1) on page 2 for the $X^{(k)}_{-}=:X_{k}$, and
equation (3.5) on page 7 for $X^{(l)}_{+}=:Y_{l}$ in \cite{Za3}}.
In our approach we work with one relatively simple $\BCH$ type
recursion, $\chi(Z)$, respectively its closed form
(\ref{chiSimple2}). Then we take the projections via $\pi_{\pm}$
to obtain $X_{\pm}(Z;t)$ up to third order in the parameter $t$.
The parameter $t$ may be interpreted as providing us with the
filtration (in the sense of formal power series). Hereby we use
heavily the relations in (\ref{rel2}). This seems to offer a
simpler way for calculating the Lie algebra elements $X_{\pm}(Z;t)
\in \frakg_{\pm}$. We only need higher expansion terms for $\chi$
and then project into $\frakg_{\pm}$. Relations (\ref{rel2})
simplify the last step considerably.


\vspace{0.5cm} {\emph{Acknowledgements}}: The first author
acknowledges greatly the support by the European Post-Doctoral
Institute and Institut des Hautes \'Etudes Scientifiques
(I.H.\'E.S.). He profited from discussions with M.~Aguiar,
J.~M.~Gracia-Bond\'{i}a and D.~Kreimer. Thanks goes to the Theory
Department at the Physics Institute of Bonn University for warm
hospitality. The second author thanks support from the NSF grant
DMS-0505643 and Rutgers University Research Council, and thanks
I.H.\'E.S. and Max Planck Institute for Mathematics in Bonn for
hospitality. Many thanks go to J.~Stasheff for comments and we
appreciate helpful discussions with K.~Barron, Y.~Huang and
J.~Lepowsky. B.~Fauser's useful remark is acknowledged. The third
author greatly acknowledges constant support from the Centre
National de la Recherche Scientifique (C.N.R.S.).



\begin{thebibliography}{abcdsfgh}

\bibitem[Agu2000]{Aguiar} M.~Aguiar,
                        {\textsl{Prepoisson algebras}},
                        Lett.~Math.~Phys., {\bf{54}}, no. 4, 263--277, (2000).

\bibitem[ABS2003]{A-B-S} M. Aguiar, N. Bergeron and F. Sottile,
                        {\textsl{Combinatorial Hopf algebras and generalized Dehn--Sommerville relations}},
                        Composito Mathematica, {\bf{142}}, 1--30, 2006.
                        [{\texttt{{\eightrm{arXiv:math.CO/0310016}}}}]

\bibitem[Atk1963]{Atkinson} F.~V.~Atkinson,
                           {\textsl{Some aspects of Baxter's functional equation}},
                           J.~Math.~Anal.~Appl., {\bf{7}}, 1--30, (1963).

\bibitem[BBT2003]{BBT} O.~Babelon, D.~Bernard and M.~Talon,
                              {\textsl{Introduction to classical integrable systems}},
                              Cambridge Monographs on Mathematical Physics.
                              Cambridge University Press, Cambridge, (2003).

\bibitem[BHL2000]{B-H-L} K.~Barron, Y.~Huang and J.~Lepowsky,
                    {\emph{Factorization of formal exponentials and uniformization}},
                    {J.~Algebra}, {\bf{228}}, 551--579, (2000).

\bibitem[Bax1960]{Baxter} G.~Baxter,
                    {\emph{An analytic problem whose solution follows from a simple algebraic identity}},
                    Pacific J.~Math., {\bf{10}}, 731--742, (1960).

\bibitem[BelDri1982]{BelavinDrinfeld1} A.~A.~Belavin and V.~G.~Drinfeld,
                            {\emph{Solutions of the classical Yang-Baxter equation for simple Lie algebras}},
                            {Funct.~Anal.~Appl.}, {\bf 16}, 159--180, (1982).

\bibitem[BoPa1957]{BoPa57} N.~N.~Bogoliubov and O.~S.~Parasiuk,
                      {\textsl{On the multiplication of causal functions in the quantum theory of fields}}.
                      Acta Math., {\bf{97}}, 227--266, (1957).

\bibitem[Car1972]{Cartier} P.~Cartier,
                    {\textsl{On the structure of free Baxter algebras}},
                    Advances in Math., {\bf{9}}, 253--265, (1972).

\bibitem[CK1999]{CK1} A.~Connes and D.~Kreimer,
                        {\textsl{Hopf algebras, Renormalization and Noncommutative Geometry}},
                        Comm.~in Math.~Phys., {\bf{199}}, 203--242, (1998).
                        [{\texttt{{\eightrm{arXiv:hep-th/9808042}}}}]

\bibitem[CK2000]{CK2} A.~Connes and D.~Kreimer,
                         {\emph{Renormalization in quantum field theory and the Riemann-Hilbert problem. I.
                         The Hopf algebra structure of graphs and the main theorem}},
                         Comm.~in Math.~Phys., {\bf{210}}, no. 1, 249--273, (2000).
                         [{\texttt{{\eightrm{arXiv:hep-th/0003188}}}}]

\bibitem[CK2001]{CK3} A.~Connes and D.~Kreimer,
                        {\textsl{Renormalization in quantum field theory and the Riemann--Hilbert problem. II.
                        The $\beta$-function, diffeomorphisms and the renormalization group}},
                        Comm.~in Math.~Phys., {\bf{216}}, 215--241, (2001).
                        [{\texttt{{\eightrm{arXiv:hep-th/0003188}}}}]

\bibitem[Col1984]{Collins84} J.~C.~Collins,
                                {\textsl{Renormalization}},
                                Cambridge Monographs on Mathematical Physics,
                                Cambridge University Press, Cambridge, (1984).

\bibitem[EGK2004]{E-G-K2} K.~Ebrahimi-Fard, L.~Guo and D.~Kreimer,
                    {\emph{Spitzer's Identity and the Algebraic Birkhoff Decomposition in pQFT}},
                    {J.~Phys.~A: Math.~Gen.}, {\bf{37}}, 11037--11052, (2004).
                    [{\texttt{{\eightrm{arXiv:hep-th/0407082}}}}]

\bibitem[EGK2005]{E-G-K3} K.~Ebrahimi-Fard, L.~Guo and D.~Kreimer,
                      {\emph{Integrable Renormalization II: the General case}},
                      Ann.~H.~Poincar{\'e}, {\bf{6}}, 369--395, (2005).
                      [{\texttt{{\eightrm{arXiv:hep-th/0403118}}}}]

\bibitem[EGGV2006]{EGGV} K.~Ebrahimi-Fard, J.~M.~Gracia-Bond\'{i}a, L.~Guo and J.C. V\'arilly,
                        {\textsl{Combinatorics of renormalization as matrix calculus}},
                        Phys.~Lett.~B., {\bf{632}}, no 4, 552--558, (2006).
                        [{\texttt{{\eightrm{arXiv:hep-th/0508154}}}}]

\bibitem[EG2005]{EG} K.~Ebrahimi-Fard and L.~Guo,
                        {\textsl{Matrix Representation of Renormalization in Perturbative Quantum Field
                        Theory}}, submitted, preprint: August 2005
                        \texttt{arXiv:hep-th/0508155}.

\bibitem[EK2005]{EK} K.~Ebrahimi-Fard and D.~Kreimer,
                        {\textsl{Hopf algebra approach to Feynman diagram calculations}},
                        J.~Phys.~A: Math.~Gen., {\bf{38}}, R385-R406, 2005.
                        [{\texttt{{\eightrm{arXiv:hep-th/0510202}}}}]

\bibitem[FGB2005]{FG} H.~Figueroa and J.~M.~Gracia-Bond\'{i}a,
                        {\textsl{Combinatorial Hopf algebras in quantum field theory I}},
                        Reviews of Mathematical Physics, {\bf{17}}, 881--976, (2005).
                        [{\texttt{{\eightrm{arXiv:hep-th/0408145}}}}]

\bibitem[Go1982]{Go} R.~Godement,
                        {\textsl{Introduction \`a la th\'eorie des groupes de Lie}},
                        Reprint of the 1982 original. Springer-Verlag, Berlin (2004).

\bibitem[Hepp1966]{Hepp66} K.~Hepp,
                           {\emph{Proof of the Bogoliubov--Parasiuk theorem on renormalization}},
                           Comm.~in Math.~Phys., {\bf{2}}, 301--326, (1966).

\bibitem[Krei1998]{KreimerHopf} D.~Kreimer,
                              {\textsl{On the Hopf algebra structure of perturbative quantum field theories}},
                              Adv.~Theor.~Math.~Phys., {\bf{2}}, 303--334, (1998).
                              [{\texttt{{\eightrm{arXiv:q-alg/9707029}}}}]

\bibitem[Krei1999]{KreimerChen} D.~Kreimer,
                        {\textsl{Chen's iterated integral represents the operator product expansion}},
                        Adv.~Theor.~Math.~Phys., {\bf{3}}, no. 3, 627--670, (1999).
                        [{\texttt{{\eightrm{{\eightrm{arXiv:hep-th/9901099}}}}}}]

\bibitem[Krei2002]{KreimerRev} D.~Kreimer,
                              {\textsl{Combinatorics of (perturbative) Quantum Field Theory}},
                              Phys.~Rep., {\bf{363}}, 387--424, (2002).
                              [{\texttt{{\eightrm{arXiv:hep-th/00110059}}}}]

\bibitem[King1962]{Kingman} J.~F.~C.~Kingman,
                    {\textsl{Spitzer's identity and its use in probability theory}},
                    J.~London~Math.~Soc., {\bf{37}}, 309--316, (1962).

\bibitem[Lod1994]{Loday94} J.-L.~Loday,
                                {\emph{S\'erie de Hausdorff, idempotents Eul\'eriens et alg\`ebres de Hopf}},
                                Expo.~Math., {\bf{12}}, 165--178, (1994).

\bibitem[Mag1954]{Mag} W.~Magnus,
                        {\textsl{On the exponential solution of differential equations for a linear operator}},
                        Comm.~Pure Appl.~Math., {\bf{7}}, 649--673, (1954).

\bibitem[Man2001]{Ma} D.~Manchon,
                {\emph{Hopf algebras, from basics to applications to renormalization}},
                Comptes-rendus des Rencontres math\'ematiques de Glanon 2001.
                [{\texttt{{\eightrm{arXiv:math.QA/0408405}}}}]

\bibitem[MQZ2000]{Za1} H.~Z.~Munthe-Kaas, G.~R.~W.~Quispel and A.~Zanna,
                {\emph{The polar decomposition of Lie groups with involutive automorphisms}},
                Technical Report, no. {\bf{191}}, Dept. of Informatics, Univ. of Bergen, Norway, (2000).

\bibitem[MQZ2001]{Za2} H.~Z.~Munthe-Kaas, G.~R.~W.~Quispel and A.~Zanna,
                {\emph{Generalized polar decompositions on Lie groups with involutive automorphisms}},
                 Found. Comput. Math.,{\bf{1}}, no. 3, 297--324, (2001).

\bibitem[Poli2002]{Polish} A.~Polishchuk,
                    {\textsl{Classical Yang-Baxter equation and the $A\sb \infty$-constraint}},
                    Adv.~Math., {\bf{168}},  no. 1, 56--95,  (2002).

\bibitem[Reu1993]{Reutenauer} C.~Reutenauer,
                                {\textsl{Free Lie Algebras}},
                                Oxford University Press, Oxford, 1993.

\bibitem[Rota1969]{Rota1} G.-C.~Rota,
                    {\emph{Baxter algebras and combinatorial identities. I, II.}},
                    {Bull.~Amer.~Math.~Soc.}, {\bf{75}}, 325--329, (1969); ibid. {\bf{75}}, 330--334, (1969).

\bibitem[RoSm1972]{RotaSmith} G.-C.~Rota and D.~Smith,
                    {\emph{Fluctuation theory and Baxter algebras}},
                    Istituto Nazionale di Alta Matematica, {\bf IX}, 179, (1972).
                    Reprinted in: ``Gian-Carlo Rota on Combinatorics:
                    Introductory papers and commentaries'', J.P.S. Kung Ed.,
                    Contemp. Mathematicians, Birkh\"auser Boston, Boston, MA, 1995.

\bibitem[Rota1995]{Rota2} G.-C.~Rota,
                    {\emph{Baxter operators, an introduction}},
                    In: ``Gian-Carlo Rota on Combinatorics, Introductory papers
                    and commentaries", J.P.S. Kung Ed., Contemp. Mathematicians,
                    Birkh\"auser Boston, Boston, MA, 1995.

\bibitem[Rota1998]{Rota3} G.-C.~Rota,
                    {\emph{Ten mathematics problems I will never solve}},
                    Invited address at the joint meeting of the
                    American Mathematical Society and the Mexican Mathematical
                    Society, Oaxaca, Mexico, December 6, 1997.
                    DMV Mittellungen, Heft {\bf{2}}, 45--52, (1998).

\bibitem[Sem1983]{STS83} M.~A.~Semenov-Tian-Shansky,
                        {\emph{What is a classical $r$-matrix?}},
                        {Funct.~Ana.~Appl.}, {\bf{17}}, no.4, 254--272, (1983).

\bibitem[Sem2000]{STS00} M.~A.~Semenov-Tian-Shansky,
                        {\emph{Integrable Systems and Factorization Problems}},
                        Lectures given at the ''Faro International Summer School on
                        Factorization and Integrable Systems" (Sept. 2000), Birkh{\"a}user 2003.
                        [{\texttt{{\eightrm{arXiv:nlin.SI/0209057}}}}]

\bibitem[Spit1956]{Spitzer} F.~Spitzer,
                    {\emph{A combinatorial lemma and its application to probability theory}},
                    {Trans.~Amer.~Math.~Soc.}, {\bf{82}}, 323--339, (1956).

\bibitem[Var1984]{Varadarajan} V.~S.~Varadarajan,
                           {\textsl{Lie Groups, Lie Algebras, and Their Representations}},
                           Springer-Verlag, (1984).

\bibitem[Wen1962]{Wendel} J.~G.~Wendel,
                            {\emph{A brief proof of a theorem of Baxter}},
                            Math.~Scand., {\bf{11}}, 107--108, (1962).

\bibitem[Zan2004]{Za3} A.~Zanna,
                {\emph{Recurrence relations and convergence theory of the generalized polar decomposition on Lie groups}},
                {Math.~Comp.}, {\bf{73}}, no. 246, 761--776, (2004).

\bibitem[Zim1969]{Zimmermann69} W.~Zimmermann,
                        {\textsl{Convergence of Bogoliubov's method of renormalization in momentum space}},
                        Comm.~in Math.~Phys., {\bf{15}}, 208--234, (1969).
\end{thebibliography}
\end{document}